\documentclass[twocolumn]{aastex62}

\received{TBA}
\revised{TBA}
\accepted{TBA}
\submitjournal{ApJ}

\shorttitle{The transient obscuring wind in MR~2251$-$178}
\shortauthors{Mao et al.}

\begin{document}

\title{Multi-wavelength observations of the obscuring wind in the radio-quiet quasar MR\,2251-178}

\correspondingauthor{Junjie Mao}
\email{jmao@tsinghua.edu.cn}

\author[0000-0001-7557-9713]{Junjie Mao}
\affiliation{Department of Astronomy, Tsinghua University, Haidian DS 100084, Beijing, People’s Republic of China}
\affiliation{SRON Netherlands Institute for Space Research, Sorbonnelaan 2, 3584 CA Utrecht, the Netherlands}

\author[0000-0002-2180-8266]{G. A. Kriss}
\affiliation{Space Telescope Science Institute, 3700 San Martin Drive, Baltimore, MD 21218, USA. Email: gak@stsci.edu}

\author{H. Landt}
\altaffiliation{Visiting Astronomer at the Infrared Telescope Facility, which is \\ operated by the University of Hawaii under contract \\ 80HQTR19D0030 with the National Aeronautics and Space \\ Administration.}
\affiliation{Centre for Extragalactic Astronomy, Department of Physics, Durham University, South Road, Durham DH1 3LE, UK. Email: hermine.landt@durham.ac.uk}

\author[0000-0002-4992-4664]{M. Mehdipour}
\affiliation{Space Telescope Science Institute, 3700 San Martin Drive, Baltimore, MD 21218, USA. Email: mmehdipour@stsci.edu}

\author[0000-0001-5540-2822]{J. S. Kaastra}
\affiliation{SRON Netherlands Institute for Space Research, Sorbonnelaan 2, 3584 CA Utrecht, the Netherlands}
\affiliation{Leiden Observatory, Leiden University, PO Box 9513, 2300 Leiden, The Netherlands}


\author{J. M. Miller}
\affiliation{Department of Astronomy, University of Michigan, 1085 South University Avenue, Ann Arbor, MI 48109-1107, USA}

\author[0000-0003-2686-9241]{D. Stern}
\affiliation{Jet Propulsion Laboratory, California Institute of Technology, Pasadena, CA 91109, USA}

\author{L. C. Gallo}
\affiliation{Department of Astronomy and Physics, Saint Mary’s University, 923 Robie Street, Halifax, NS, B3H 3C3, Canada}

\author[0000-0003-3678-5033]{A. G. Gonzalez}
\affiliation{Department of Astronomy and Physics, Saint Mary’s University, 923 Robie Street, Halifax, NS, B3H 3C3, Canada}

\author{J. J. Simon}
\affiliation{Jet Propulsion Laboratory, California Institute of Technology, Pasadena, CA 91109, USA}

\author{S. G. Djorgovski}
\affiliation{Division of Physics, Mathematics, and Astronomy, California Institute of Technology, Pasadena, CA 91125, USA}

\author{S. Anand}
\affiliation{Division of Physics, Mathematics, and Astronomy, California Institute of Technology, Pasadena, CA 91125, USA}

\author[0000-0002-5619-4938]{Mansi M. Kasliwal}
\affiliation{Division of Physics, Mathematics, and Astronomy, California Institute of Technology, Pasadena, CA 91125, USA}

\author{V. Karambelkar}
\affiliation{Division of Physics, Mathematics, and Astronomy, California Institute of Technology, Pasadena, CA 91125, USA}



\begin{abstract}
Obscuring winds driven away from active supermassive black holes are rarely seen due to their transient nature. They have been observed with multi-wavelength observations in a few Seyfert 1 galaxies and one broad absorption line radio-quiet quasar so far. An X-ray obscuration event in MR\,2251-178 was caught in late 2020, which triggered multi-wavelength (NIR to X-ray) observations targeting this radio-quiet quasar. In the X-ray band, the obscurer leads to a flux drop in the soft X-ray band from late 2020 to early 2021. X-ray obscuration events might have a quasi-period of two decades considering earlier events in 1980 and 1996. In the UV band, a forest of weak blueshifted absorption features emerged in the blue wing of Ly$\alpha$ $\lambda1216$ in late 2020. Our XMM-Newton, NuSTAR, and HST/COS observations are obtained simultaneously, hence, the transient X-ray obscuration event is expected to account for the UV outflow, although they are not necessarily caused by the same part of the wind. Both blueshifted and redshifted absorption features were found for He {\sc i} $\lambda10830$, but no previous NIR spectra are available for comparison. The X-ray observational features of MR\,2251-178 shared similarities with some other type 1 AGNs with obscuring wind. However, observational features in the UV to NIR bands are distinctly different from those seen in other AGN with obscuring winds. A general understanding of the observational variety and the nature of obscuring wind is still lacking.
\end{abstract}

\keywords{black hole physics -- quasars: individual (MR\,2251-178) -- X-rays, ultraviolet, infrared: galaxies -- techniques: spectroscopic}



\section{Introduction}
\label{sct:intro}
At the center of almost all galaxies, there is a supermassive black hole \citep{Netzer2015}. Active Galactic Nuclei (AGN) are the observed manifestation of the inflow of matter onto supermassive black holes. The textbook anatomy of an AGN not only includes a supermassive black hole fed by an accretion disk but also clouds in the broad- and narrow-line regions, as well as the dusty torus \citep{Antonucci1993}. Clouds in the broad- and narrow-line regions account for the broad (with a typical velocity width $\gtrsim1000~{\rm km~s^{-1}}$) and narrow ($\lesssim1000~{\rm km~s^{-1}}$) emission lines observed in the optical spectra \citep{Padovani2017}.

Ionized winds driven away from black holes have also been observed \citep{Crenshaw2003,Tombesi2013,Kaastra2014}, which might be another key element to be included in the textbook anatomy of AGN \citep{Antonucci1993}. These ionized winds are thought to play an important role in the evolution of black holes and their host galaxies \citep{Laha2021}. In the X-ray band, three types of ionized winds have been observed so far: warm absorbers, ultrafast outflows, and obscurers. The classical warm absorbers are identified with multiple narrow absorption lines with a typical outflow velocity of $\lesssim 10^3~{\rm km~s^{-1}}$ \citep{Crenshaw2003}. Ultrafast outflows are mainly inferred from the absorption features of highly ionized Fe {\sc xxvi} and/or Fe {\sc xxv} in the hard X-ray band \citep[e.g.,][]{Reeves2003,Tombesi2013,Parker2017}. The outflow velocity of ultrafast outflows can reach up to about a third of the speed of light ($\sim10^{4-5}~{\rm km~s^{-1}}$). Obscuring winds have outflow velocities up to $\sim6000~{\rm km~s^{-1}}$ \citep{Kaastra2014,Mehdipour2017}, larger than that of the warm absorber but smaller than that of ultrafast outflows. The outflow velocities are currently measured through discrete blueshifted absorption features in the UV band during the soft X-ray obscuration period, where the soft X-ray flux are significantly lowered with no discrete absorption features observable with current instruments \citep{Mao2022}. Furthermore, ultrafast outflows occupy the high column density and ionization parameter ($N_{\rm H}-\xi$) part of the parameter space, while warm absorbers occupy the other side of the parameter space. Obscuring winds are in between, overlapping more with the warm absorber.

In the past few years, transient obscuring winds have been reported with multi-wavelength observations in a few nearby ($z\lesssim0.16$) Seyfert galaxies: NGC\,5548 \citep{Kaastra2014}, NGC\,985 \citep{Ebrero2016}, NGC\,3783 \citep{Mehdipour2017,Kaastra2018a}, Mrk\,335 \citep{Longinotti2013, Longinotti2019}, Mrk\,817 \citep{Miller2021,Kara2021}, and NGC\,3227 \citep{Mehdipour2021,Mao2022}. In addition, obscuring winds were also inferred from archival joint X-ray and UV observations of the broad absorption line (BAL) quasar PG 2112+059 at $z\sim0.46$ \citep{Saez2021}. The duration of the obscuration varies from hours \citep[e.g., NGC\,3227,][]{YJWang2022} to years \citep[e.g., NGC\,5548,][]{Mehdipour2016a}. During the obscuration period, the wind properties (e.g., the line-of-sight hydrogen column density, ionization parameter, and covering factor) can also change over time \citep{DeRosa2015,Mehdipour2016a,Cappi2016}. In UV grating spectra, the obscurers might leave their fingerprints as blueshifted broad absorption troughs in the blue wing of prominent broad emission lines like Ly$\alpha$ $\lambda1216$ and C {\sc iv} $\lambda\lambda1548,1550$ \citep[e.g.,][]{Kriss2019a}. For the long-lasting obscurer in NGC\,5548, a blueshifted broad absorption trough of He {\sc i}* $\lambda10830$ was observed in the NIR \citep{Landt2019, Wildy2021}. The obscurers are also expected to screen photons from the central engine to the distant narrow-line region \citep{Kaastra2014,Mehdipour2017}. This shielding effect \citep{Proga2004} can give rise to the emergence of low-ionization narrow absorption lines in the UV and NIR bands \citep{Arav2015,Kriss2019a,Wildy2021}.

Obscurers might contribute to the weakening and disappearance of broad emission lines if the obscuration effect is not limited to our line-of-sight. For NGC\,5548, a ``broad-line holiday" (from April to July 2014) was discovered \citep{Goad2016}, where broad emission lines of highly-ionized species like C {\sc iv} $\lambda\lambda1548,1550$ decorrelate from the variations of the FUV continuum. From the theoretical perspective, \citet{Dehghanian2019a, Dehghanian2019b} suggest that a dense wind with its base extended down to the accretion disk might provide the physical explanation of the ``broad-line holiday", as well as the aforementioned UV and X-ray spectral features observed in NGC\,5548. Moreover, if the wind is adequately dense, even the ``changing-look" behavior of some AGN might be explained \citep{Dehghanian2019b}. For Mrk\,817, in a period of 55 days in late 2020, when the target was heavily obscured, the UV continuum and the UV broad emission line variability decoupled \citep{Kara2021}.

MR\,2251-178 is the X-ray brightest radio-quiet quasar \citep{Cooke1978, Ricker1978}. It has a black hole mass of $(2.0\pm0.5)\times10^8~M_{\odot}$ \citep{Lira2011} and is at $z=0.06398$ \citep{Beckmann2006}. By comparing two X-ray spectra observed with the Einstein Observatory in 1979 and 1980, \citet{Halpern1984} noticed an increase in the absorbing column density by a large factor over the $\sim1$~yr interval. Using Rossi X-ray Timing Explorer (RXTE) data from 1996 to 2012, \citet{Markowitz2014} also identified a clear obscuration event in 1996. Here, we present another obscuration event in MR\,2251-178 in late 2020 caught with multi-wavelength observations, including a Swift monitoring campaign, coordinated observations with Hubble Space Telescope (HST), XMM-Newton, and NuSTAR on 2020-12-16, and NIR and optical spectroscopy using NASA's Infrared Telescope Facility, Gemini, Keck, and Palomar from September to December 2020.

\begin{deluxetable*}{c|c|c}
\tablecaption{Observation log of MR\,2251-178 used in the present work. All archival Swift observations were also used, but are not listed here. The Swift target IDs are 00049534, 00092238, 00093126, 00093159, 00094001, 00095001, 00095654, 00081592, and 00089190.
\label{tbl:obs_log}}
\tablehead{
\colhead{Observatory} & \colhead{Date} &\colhead{Note (e.g., ObsIDs and duration) }
}
\startdata
Palomar/TriSpec & 2020-12-22 & $\sim9700-24600$~\AA\ (0.3 ks) \\
NuSTAR & 2020-12-16 &  90601637002 (25 ks) \\
XMM-Newton & 2020-12-16 & 0872390801 (50 ks) \\
HST/COS & 2020-12-16 & LEHV010 (1.2 ks) \\
Palomar/DBSP & 2020-12-11 & $\sim3000-7000$~\AA\ (0.3 ks)\\
Keck/LRIS & 2020-10-18 & $\sim3200-5600$~\AA\ + $5542-10347$~\AA\ (0.3 ks) \\
Gemini/GNIRS & 2020-09-30 & GN-2020B-FT-111, $8200-25148$~\AA\ (3.6 ks) \\
IRTF/SpeX & 2020-09-06 & $\sim6900-25680$~\AA (7.2 ks) \\
\noalign{\smallskip}
HST/COS & 2011-09-29 & LBGB030 (10.0 ks) \\
XMM-Newton & 2002-05-18 & 0012940101 (65~ks) \\
FUSE & 2001-06-20 & P1111010000 (52 ks) \\
HST/FOS & 1996-08-02 & Y3AI200 (6.8~ks) \\
\noalign{\smallskip}
\enddata
\end{deluxetable*}

\section{Observations and data reduction}
\label{sct:obs_dr}
Apart from the archival Swift observations from 2014-01-09 to 2021-05-10, NIR to X-ray observations used in the present work are provided in Table~\ref{tbl:obs_log}. In the same table, some archival observations are also included for comparison purposes. Note that MR\,2251-178 was put in the spotlight by Chandra (in coordination with HST/COS) and XMM-Newton (about one month after the Chandra observations) in 2011. The high-quality X-ray data were analyzed extensively in \citet[][for Chandra]{Reeves2013} and \citet[][for XMM-Newton]{Nardini2014}. We do not include these observations in this work because our main focus is the transient obscuring wind observed in late 2020 with multi-wavelength observing facilities. We do not find obscuring winds in the 2011 X-ray (and UV) data, where the broadband continuum level was much higher than those in 2002 and 2020.

\subsection{Swift}
The Neil Gehrels Swift Observatory \citep{Gehrels2004} provides the most efficient way to track the long-term variation of the X-ray and UV flux of AGN. The X-ray flux was provided by the X-ray telescope \citep[XRT,][]{Burrows2005} operating in imaging photon counting (PC) mode. The soft ($0.3-1.5$~keV) and hard ($1.5-10$~keV) band flux were obtained with the online XRT pipeline \citep{Evans2009}. The UV flux was covered by the UltraViolet and Optical Telescope \citep[UVOT,][]{Roming2005}. The UVW2 (centered around 1928~\AA) flux, which historically has shown the strongest UV variability, was used here. The \textit{uvotsource} tool was used to perform aperture photometry.

The obscurer can absorb a large fraction of the soft X-ray flux, thus, leading to an elevated X-ray hardness ratio. Simultaneous UV and X-ray data enable us to distinguish variability caused by an obscurer from that of the AGN continuum \citep[e.g.,][]{Mehdipour2016a, Mehdipour2017, Mehdipour2022a}.

\begin{figure*}
\centering
\includegraphics[width=.83\hsize, trim={0.cm 0.5cm 2.cm 1.0cm}, clip]{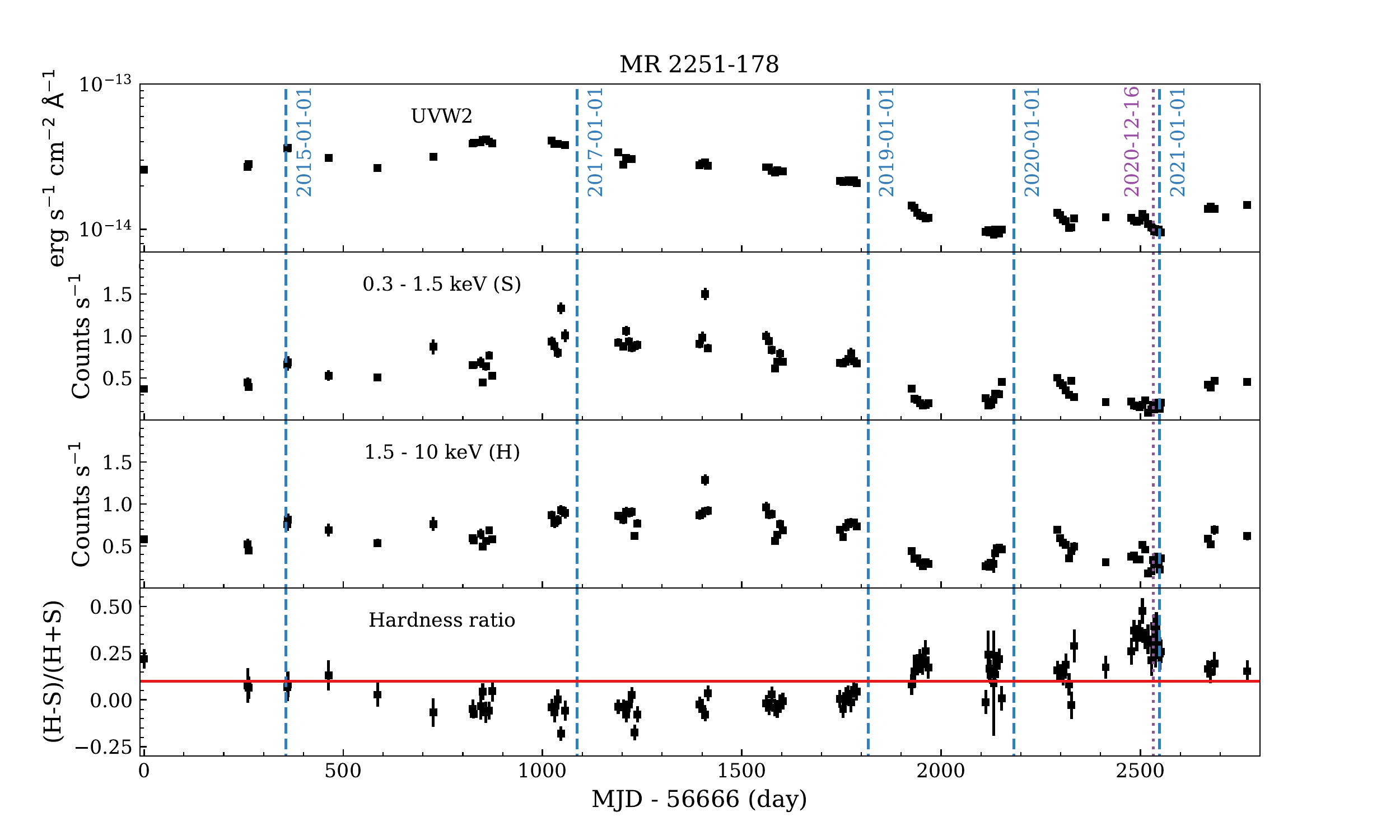}
\caption{Archival Swift XRT and UVOT observations of MR\,2251-178 from 2014-01-09 to 2021-05-10. The top panel is the UVW2 flux. The two middle panels are the count rates in the hard (H: $1.5-10$~keV) and soft (S: $0.3-1.5$~keV) X-ray bands. The bottom panel shows the X-ray hardness ratio $R=(H-S)/(H+S)$. The horizontal solid line in red is the historical average hardness ratio before late 2020. The vertical dashed line in purple marks the joint XMM-Newton + HST/COS + NuSTAR observation on 2020-12-16.}
\label{fig:plot_ltc_uvx}
\end{figure*}

\begin{figure}
\centering
\includegraphics[width=\hsize, trim={0.cm 0.cm 0.5cm 0.2cm}, clip]{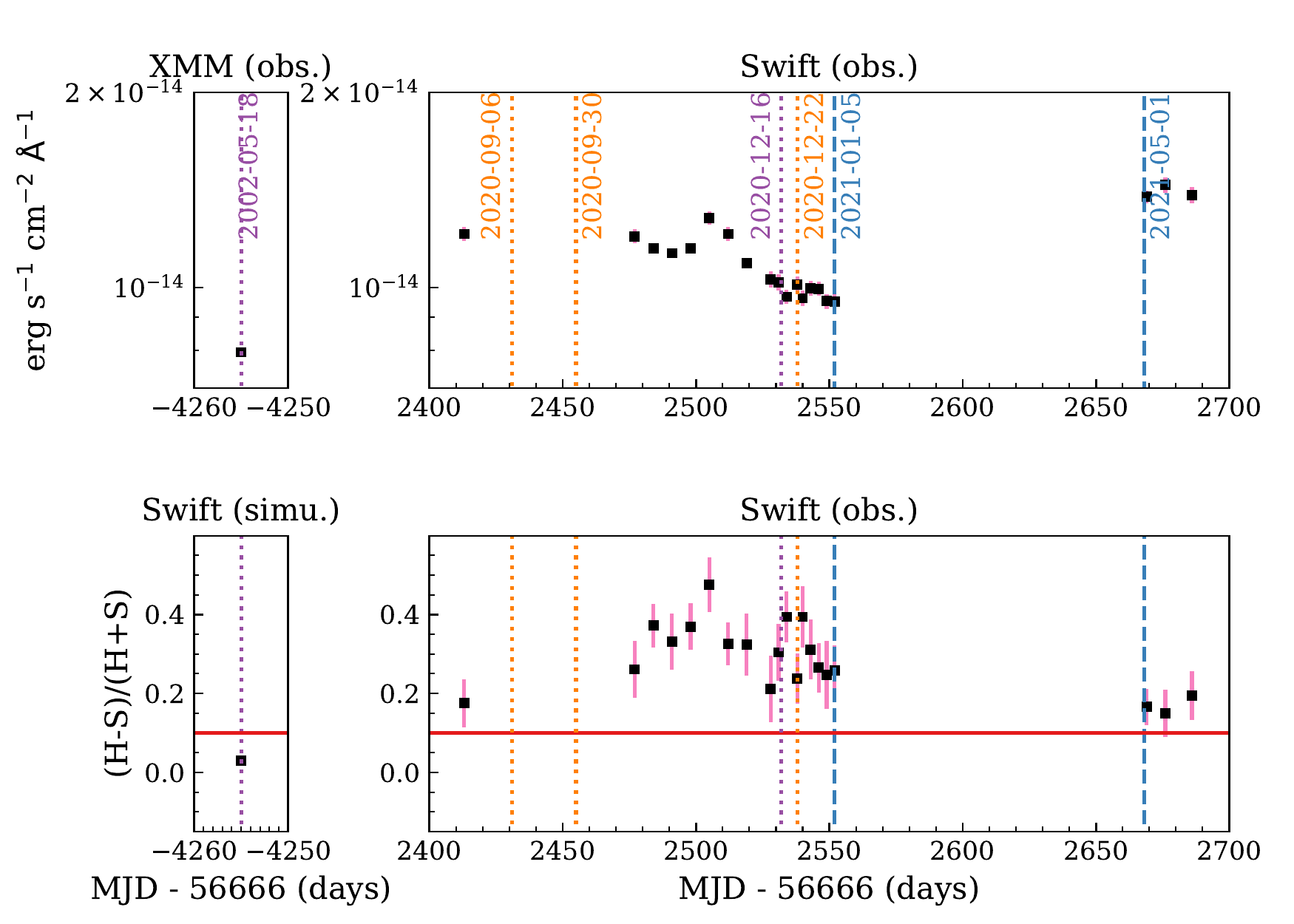}
\caption{A close-up view of the Swift UVW flux and X-ray hardness ratio from Aug. 2020 to May. 2021, as well as in 2002. The uncertainties of the observed UVW2 flux are too small to be seen in the plotting scale. The 2002 UVW2 flux is obtained with the XMM-Newton Optical Monitor, while the Swift X-ray hardness ratio is simulated based on the best-fit model to this XMM-Newton observation. }
\label{fig:plot_uv_xhr}
\end{figure}

For MR\,2251-178, archival Swift observations date back to 2014-01-09 (Figure~\ref{fig:plot_ltc_uvx}). Since the summer of 2020, the Swift X-ray hardness ratio has been gradually increasing (Figure~\ref{fig:plot_uv_xhr}). Therefore, we requested multi-wavelength spectroscopic observations of MR\,2251-178 with facilities on the ground and in space from September to December 2020, including a coordinated campaign with XMM-Newton, HST/COS, and NuSTAR  on 2020-12-16, to confirm the obscuration event and to study the nature of the obscuring material. The Swift X-ray hardness ratio of MR\,2251-178 remained at a relatively high level up to 2021-01-05. The target was out of visibility between January and April 2021. In May 2021, the hardness ratios returned to the historical average value.

\subsection{XMM-Newton}
XMM-Newton \citep{Jansen2001} observations provide high-quality X-ray spectra in the energy band of $0.3-10$~keV. The Reflection Grating Spectrometer \citep[RGS,][]{denHerder2001} provides high-resolution soft X-ray spectra below $\sim$2~keV, which can reveal discrete emission and absorption features. The positive-negative junction (pn) CCD (charge-coupled device) camera of the European Photon Imaging Camera (EPIC) instrument can cover hard X-ray spectra in the $2-10$ keV energy band. Adding EPIC/MOS data does not significantly increase the statistics and will complicate the analysis due to cross-calibration among different instruments. We used the XMM-Newton Science Analysis Software (SAS) v19.0 for data reduction following standard procedures. We also include an XMM-Newton observation of MR\,2251-178 obtained on 2002-05-18 (observation ID: 0012940101) for comparison purposes. Both $7-38$~\AA\ RGS data and $1.55-10$~keV EPIC/pn data were used for spectral analysis (Sect.~\ref{sct:mo}). To correct for the cross-calibration between RGS and EPIC-pn spectra, the latter were re-scaled by a factor of 0.950 (2002) and 0.966 (2020) to match the flux level in the $7-8$~\AA\ overlapping wavelength range.

\subsection{NuSTAR}
NuSTAR \citep{Harrison2013} complements X-ray spectra obtained with XMM-Newton in the energy band up to $79$~keV. Our NuSTAR data were reduced with the NuSTAR Data Analysis Software (NUSTARDAS) and CALDB version 20201101. The FPMA and FPMB source spectra were extracted from a circular region centered on the point source with a radius of 80 arcsec. The background spectra of equivalent extraction radius were extracted from a nearby source-free region. The
$5-78$~keV NuSTAR data were used for spectral analysis (Sect.~\ref{sct:mo}). To correct for the cross-calibration among spectra collected by XMM-Newton and NuSTAR, the latter were re-scaled (with respect to RGS) by a factor of 0.962 (FPMA) and 1.045 (FMPB) to match the flux level in the $5-10$~keV overlapping energy range.

\subsection{HST/COS}
The Cosmic Origin Spectrograph aboard the Hubble Space Telescope \citep[HST/COS,][]{Green2012} provides high-quality FUV spectra in the wavelength range $1070-1800$~\AA\ at a resolving power of $R\sim$15000. We use both the archival observations in 2011 and the new single-orbit observation in 2020. The archival observations used grating G130M with central wavelength settings of 1291~\AA, 1300~\AA, 1309~\AA, and 1318~\AA, and G160M with central wavelengths settings of 1589~\AA, 1600~\AA, 1611~\AA, and 1623~\AA. In 2020, we used G130M at the 1222~\AA\ central wavelength setting and all four FPPOS locations along with G160M at central wavelengths settings of 1533~\AA\ and 1589~\AA. The multiple central wavelengths and FPPOS settings permit our spectra to span the gaps between detector segments and enable us to eliminate detector artifacts and improve the flat-field properties by sampling each wavelength at different detector locations. We retrieved calibrated data from the Mikulski Archive for Space Telescopes (MAST) and reprocessed with COS pipeline version 3.3.10, which incorporates all improvements related to wavelength scales, error propagation, and time-dependent sensitivity. Measurement of the centroids of interstellar absorption features revealed no need for any zero-point corrections to the wavelength scale, which is accurate to 5 $\rm km~s^{-1}$ \citep{Dashtamirova2020}.

To extend the archival 2011 spectrum to shorter wavelengths, we used FUSE observations from 2001 with optimal reprocessing as described by \cite{Kriss2006}, and scaled the flux of the FUSE spectrum to match that of COS in the overlapping wavelength range of $1135-1180$~\AA. An archival HST Faint Object Spectrograph (FOS) observation from 1996 was also used for comparison purposes.

\subsection{NIR and optical spectroscopy}
We obtained NIR spectroscopy on three occasions. To our knowledge, these data are the first NIR spectra of MR~2251$-$178. On 2020-09-06, we used the SpeX instrument \citep{Rayner2003} on NASA's Infrared Telescope Facility (IRTF) equipped with the short cross-dispersed mode (SXD, $0.7-2.55~\mu$m) and a $0.3'' \times 15''$ slit, which we oriented at the parallactic angle. This setup gives an average spectral resolving power of $R=2000$. The on-source exposure time was $60\times120$~s and the weather was photometric with a seeing of $0.46$ arcsec, resulting in a high-quality spectrum with an average continuum $S/N \sim 60$. Before the science target, we observed the nearby (in position and airmass) A0~V star HD~218639 and used this standard star to correct our science spectrum for telluric absorption and for flux calibration. We reduced the data using Spextool (version 4.1), an Interactive Data Language (IDL)-based software package developed for SpeX \citep{Cushing2004} which carries out all the procedures necessary to produce fully reduced spectra.

On 2020-09-30, we obtained a similar cross-dispersed NIR spectrum with the Gemini Near-Infrared Spectrograph \citep[GNIRS;][]{gnirs} at Gemini North in queue mode (Program ID: GN-2020B-FT-111). We chose a slit of $0.3'' \times 7''$, which we oriented at the parallactic angle, and obtained an on-source exposure time of $16 \times 120$~s. This setup resulted in a spectrum with an average resolution of $R=1400$ and a continuum $S/N \sim 100$. We reduced the data using the Gemini/IRAF package (version 1.13) with GNIRS specific tools \citep{gnirssoft}. We again selected HD~218639 as our standard star for telluric correction and flux calibration.

Then, one week after the coordinated XMM-Newton, HST/COS, and NuSTAR observations, we obtained on 2020-12-23 a final cross-dispersed NIR spectrum, with TripleSpec \citep{Wilson2004} at the Palomar~5~m. The entrance slit for this instrument is fixed at $1'' \times 30''$, which gives a spectral resolving power of $R=2700$. The weather was good and the on-source exposure time of $32 \times 120$~s resulted in a spectrum with a continuum $S/N \sim 15$. The somewhat reduced quality of this spectrum relative to those from the IRTF and Gemini is due to a high airmass ($\sec z=1.782$) and a higher spectral resolution.

We obtained optical spectroscopy on two occasions. On 2020-10-18, we used the Low Resolution Imaging Spectrometer \citep[LRIS;][]{lris} mounted on the Keck~10~m telescope equipped with the 600/4000 and 400/8500 gratings for the blue and red arms, respectively, and the $1.5''$~slit. This setup gives a relatively large spectral coverage of $\sim 3100 - 10300$~\AA, with a very small spectral gap of $\sim 40$~\AA~between the two arms. The average spectral resolving power is $R=750$. The slit was rotated to the parallactic angle, but note that LRIS has an atmospheric dispersion corrector. The on-source exposure time was $300$~s at an average airmass of $\sec z=1.349$, and resulted in an average continuum $S/N \sim 70$. The data were reduced using standard longslit routines from the IRAF software package.

Five days before the coordinated multi-wavelength observations, we obtained on 2020-12-11 an optical spectrum at the Palomar~5~m with the Double Spectrograph (DBSP) instrument\citep{dbsp}. Similar to LRIS, a dichroic splits the light into separate blue and red channels and we chose for these the 600/4000 and 316/7500 gratings, respectively. We observed through a $2''$~slit oriented at a parallactic angle for a total exposure time of $300$~s. This setup resulted in a spectrum with an average resolving power of $R=750$ and a continuum $S/N \sim 25$. The night was not photometric, hence, the entire DBSP spectrum needs to be increased by a factor of 1.7 to match the O [{\sc iii}] $\lambda5007$ line peak of Keck/LRIS.

\section{Spectral analysis}
\label{sct:mo}
We use SPEX v3.05.00 \citep{Kaastra1996,Kaastra2018b} and $C$-statistics for the X-ray spectral analysis \citep{Kaastra2017}. Statistical uncertainties are quoted at the 68\% confidence level. The protosolar abundances of \citet{Lodders2009} were used for all the plasma models.

We include model components described as the following to account for continuum, emission, and absorption features simultaneously. For the intrinsic continuum, the NIR to UV data collected in 2020 provide constraints above the Lyman limit ($\lambda>912$~\AA), while the X-ray data provide constraints below the Lyman limit.

\subsection{Spectral model components above the Lyman limit}
\label{sct:cont_nir2uv}
The AGN continuum above the Lyman limit can be described as a disk blackbody model. We use the \textit{dbb} in SPEX, which is a geometrically thin but optically thick Shakura-Sunyaev accretion disk \citep{Shakura1973}. Its normalization and temperature are left free during the fitting. The Milky Way reddening along our line-of-sight is taken into account with frozen parameters $E(B-V)=0.039$ \citep{Schlegel1998} and $R_V=3.1$, as used by \citet{Nardini2014}.

We use line-free zones in both the HST/COS spectrum ($\sim1135-1205$, $1356-1364$, and $1428-1442$~\AA) obtained on 2020-12-16 and the Keck/LRIS spectrum ($\sim6000-6050$, $6530-6580$, $7350-7400$, and $7950-8050$~\AA) obtained on 2020-10-18 to constrain \textit{dbb} parameters. To account for the host galaxy emission in the Keck spectrum, a template starlight emission from the bulge \citep{Kinney1996} was included as a \textit{file} model with its normalization free to vary \citep{Mehdipour2015}. The modeled luminosity of the host galaxy emission is $1.47\times10^{43}~{\rm erg~s^{-1}}$ in the $1000-10000$~\AA\ wavelength range. The dusty torus (NIR excess in Figure~\ref{fig:plot_cf_sed_nir2xray}), Balmer continuum and blended Fe {\sc ii} emission (the blue bump near UV) are not included in our modeling.

\begin{figure}
\centering
\includegraphics[width=\hsize, trim={0.5cm 0.cm 1.cm .cm}, clip]{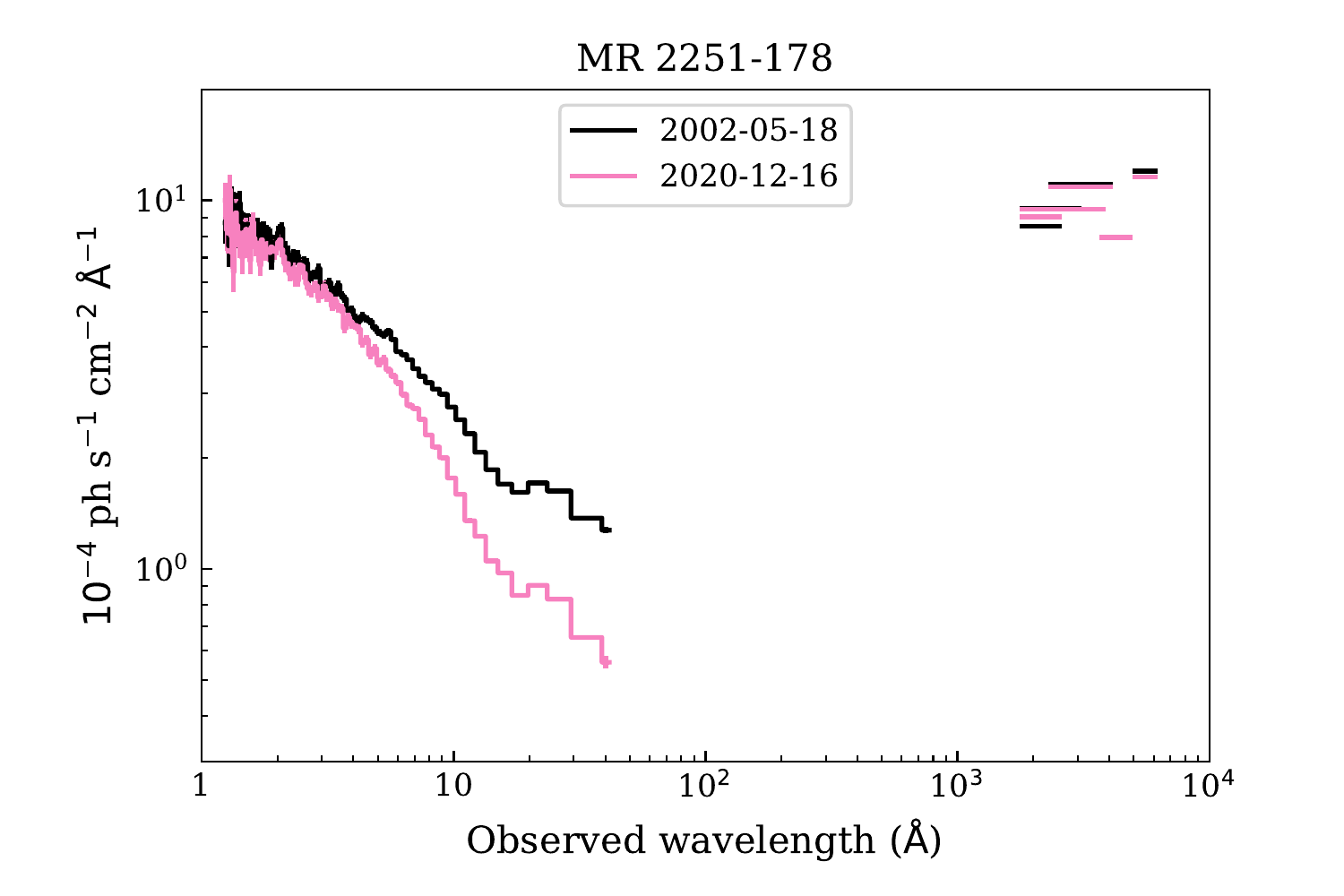}
\caption{XMM-Newton spectra of MR\,2251-178 from 2002-05-18 (black) and 2020-12-16 (pink). The 2002 observation used the UVW2, UVM2, UVW1, and V filters, while the 2020 observation used all six optical and UV filters. The apparent lowering of the soft X-ray flux in 2020 is due to the presence of the obscurer. }
\label{fig:plot_cf_spec_02vs20}
\end{figure}

\subsection{Spectral model components below the Lyman limit}
\label{sct:cont_xray}

To model the AGN continuum below the Lyman limit, we use a Comptonized disk component \citep[\textit{comt},][]{Titarchuk1994}, a power-law component, and a neutral reflection component \citep[\textit{refl},][]{Magdziarz1995,Zycki1999}. The warm Comptonization (\textit{comt}) is one of the possible interpretations of the soft X-ray excess in AGN \citep{Crummy2006,Dauser2010,Done2012,Petrucci2013}. As shown in our previous works \citep[e.g.,][]{Mehdipour2015,Mehdipour2017,Mehdipour2021}, it suits our purpose of building a broadband SED for photoionization modeling. The power-law component was cut-off exponentially below the Lyman limit and above 100 keV \citep{Orr2001}. We refer readers to A. Gonzalez et al. (in prep) for a detailed study on the X-ray continuum and its variability over the past two decades. The Galactic absorption in the X-ray band was modeled with a \textit{hot} model in SPEX. The line-of-sight hydrogen (H {\sc i} and ${\rm H_2}$) column density was fixed to $2.71\times10^{20}~{\rm cm^{-2}}$ \citep{Willingale2013}.

The photoionization model \textit{pion} \citep{Miller2015,Mehdipour2016b,Mao2017} in SPEX was used for both the warm absorber and emitter \citep[e.g.,][]{Mao2018,Mao2019}. Similar to \citet{Reeves2013}, three \textit{pion} absorption components are required with distinct parameters (hydrogen column density, ionization parameter, and kinematics) for the warm absorber. For the warm emitter, two \textit{pion} emission components are required.

As shown in Figure~\ref{fig:plot_cf_spec_02vs20}, while the flux in the optical, UV, and hard X-ray bands in 2020 are comparable to those in 2002, the soft X-ray flux was significantly lower due to the presence of the obscurer. The obscurer was modeled as an additional \textit{pion} absorption component for the 2020 data set.

\section{Results}
\label{sct:res}
\subsection{Broadband AGN SED}
\label{sct:sed}
\begin{figure*}
\centering
\includegraphics[width=.83\hsize, trim={0.5cm 0.cm 2.cm 0cm}, clip]{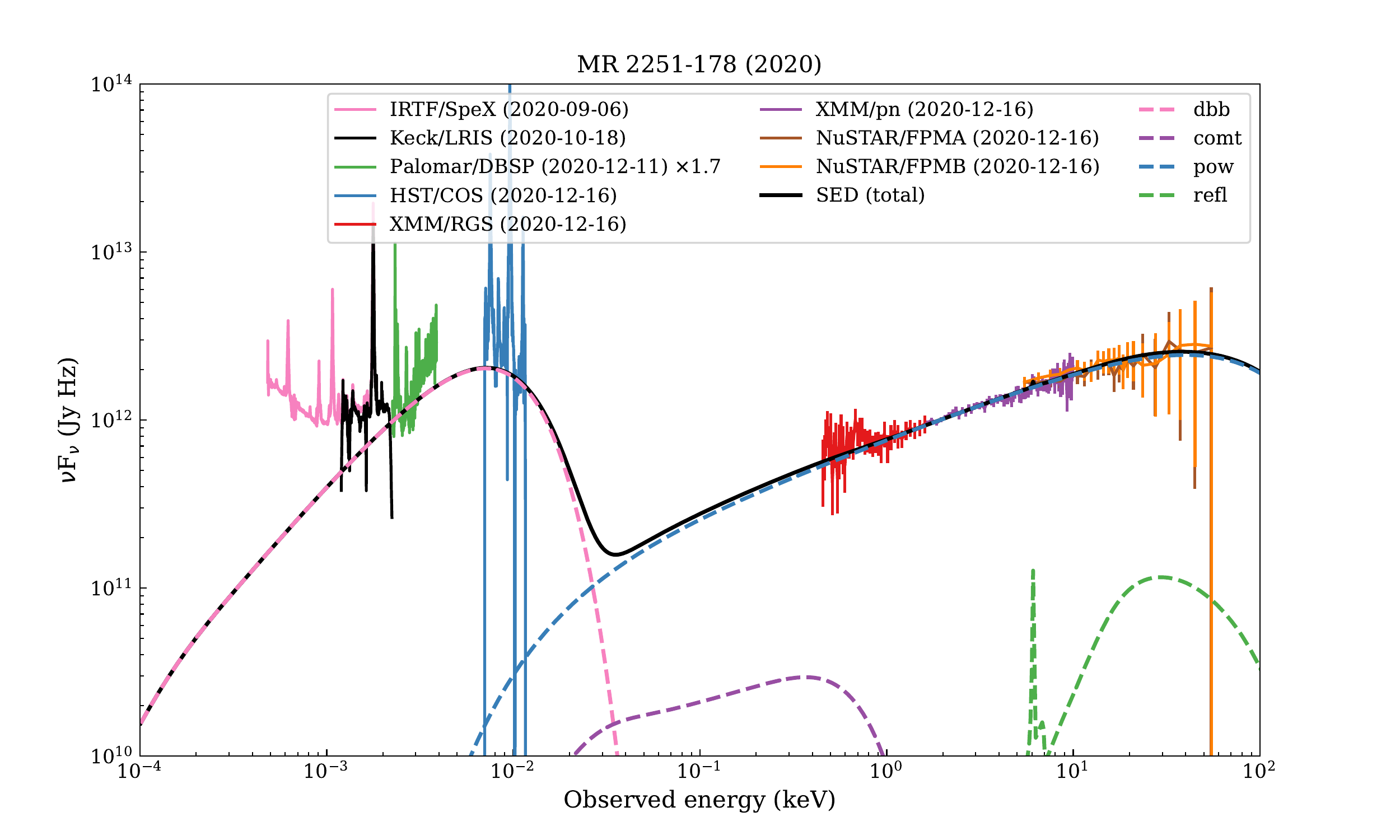}
\caption{Broadband SED model for the AGN continuum, which consists of a disk blackbody (dbb), a Comptonized disk component (comt), a power-law component (pow), and a reflection component (refl). NIR, UV, and X-ray spectra obtained with ground and space missions at different times are also shown. Both the model and data are corrected for the extinction (NIR and UV) and absorption (X-ray) effects. Observed data include emissions from warm emitter, BLR, dusty torus, and host galaxy, in addition to the intrinsic AGN SED. Data and model are rebinned for clarity. }
\label{fig:plot_cf_sed_nir2xray}
\end{figure*}

After taking into account all the obscuration, absorption, emission, and extinction effects, we derive the best-fit parameters of the broadband SED parameters for both 2002 and 2020 (Table~\ref{tbl:fit_02vs20sed}). We used relatively simple spectral model components to construct the SED, which is sufficient for the purpose of photoionization modeling. In Figure~\ref{fig:plot_cf_sed_nir2xray}, we illustrate the derived broadband 2020 SED model from NIR to X-ray.

\begin{deluxetable}{c|ccc}
\tablecaption{Best-fit parameters of the broadband spectra of MR\,2251-178 on 2002-05-18 and 2020-12-16. The C-statistics refer to the final best-fit, where all obscuration, absorption, and emission components are taken into account. All quoted errors (including the upper limits) refer to the statistical uncertainties at the 68.3\% confidence level. Frozen and coupled parameters are indicated with (f) and (c), respectively.
\label{tbl:fit_02vs20sed}}
\tablehead{
\colhead{Obs. date} & \colhead{2002-05-18} & \colhead{2020-12-16}
}
\startdata
\noalign{\smallskip}
\multicolumn{3}{c}{Disk blackbody} \\
\noalign{\smallskip}
\hline
\noalign{\smallskip}
Norm (${\rm cm^{-2}}$) & $6.2\times10^{28}$ (f) & $6.2\times10^{28}$ (f) \\
$T$ (eV) & $6.27$ (f) & $6.27$ (f) \\
\noalign{\smallskip}
\hline
\noalign{\smallskip}
\multicolumn{3}{c}{Comptonisation} \\
\noalign{\smallskip}
\hline
\noalign{\smallskip}
Norm (${\rm ph~s^{-1}~keV^{-1}}$) & $5.4_{-2.0}^{+0.5}\times10^{54}$ & $(1.3\pm0.4)\times10^{54}$ \\
$T_{\rm seed}$ (eV) & $6.27$ (c) & $6.27$ (c) \\
$T_{\rm c}$ (keV) & $0.168_{-0.004}^{+0.017}$  & $0.168$ (f)  \\
$\tau$ & $30$ (f) & $30$ (f) \\
\noalign{\smallskip}
\hline
\noalign{\smallskip}
\multicolumn{3}{c}{Power-law} \\
\noalign{\smallskip}
\hline
\noalign{\smallskip}
Norm (${\rm ph~s^{-1}~keV^{-1}}$) & $4.10_{-0.07}^{+0.11}\times10^{52}$ & $(4.5\pm0.2)\times10^{52}$ \\
$\Gamma$ & $1.522\pm0.012$ & $1.56\pm0.02$ \\
\noalign{\smallskip}
\hline
\noalign{\smallskip}
\multicolumn{3}{c}{Reflection} \\
\noalign{\smallskip}
\hline
\noalign{\smallskip}
Norm (${\rm ph~s^{-1}~keV^{-1}}$) & $4.10\times10^{52}$ (f) & $4.5\times10^{52}$ (f) \\
$\Gamma$ & $1.52$ (f) & $1.57$ (f) \\
$\log \xi~({\rm erg~s^{-1}~cm})$ & 0.0 (f) & 0.0 (f) \\
scale & $0.23_{-0.04}^{+0.04}$ & $<0.05$ \\
\noalign{\smallskip}
\hline
\noalign{\smallskip}
\multicolumn{3}{c}{Luminosity} \\
\noalign{\smallskip}
\hline
\noalign{\smallskip}
$L_{\rm 0.3-2~keV}~{\rm (erg~s^{-1})}$ & $1.6\times10^{44}$ & $1.3\times10^{44}$ \\
$L_{\rm 2-10~keV}~{\rm (erg~s^{-1})}$ & $2.2\times10^{44}$ & $2.2\times10^{44}$ \\
$L_{\rm 0.001-10~keV}~{\rm (erg~s^{-1})}$ & $8.9\times10^{44}$ & $8.2\times10^{44}$  \\
$L_{\rm 1-1000~Ryd}~{\rm (erg~s^{-1})}$ & $6.0\times10^{44}$ & $5.3\times10^{44}$  \\
\noalign{\smallskip}
\hline
\noalign{\smallskip}
\multicolumn{3}{c}{Statistics} \\
\noalign{\smallskip}
\hline
\noalign{\smallskip}
$C_{\rm stat}$~(total) & 1806.0 & 2902.4 \\
$C_{\rm expt}$~(total) & $1727\pm59$ & $2460\pm71$ \\
d.o.f. (total) & 1676 & 2335 \\
$C_{\rm stat}/C_{\rm expt}$~(RGS) & 1614.2/1537 & 1724.3/1587  \\
$C_{\rm stat}/C_{\rm expt}$~(pn) & 191.8/189 & 190.1/170  \\
$C_{\rm stat}/C_{\rm expt}$~(FPMA) & -- -- & 544.0/352  \\
$C_{\rm stat}/C_{\rm expt}$~(FPMB) & -- -- & 444.0/352 \\
\noalign{\smallskip}
\enddata
\end{deluxetable}

The \textit{dbb} parameters in Table~\ref{tbl:fit_02vs20sed} are obtained with NIR to UV data as described in Section~\ref{sct:cont_nir2uv}. They are kept frozen for the X-ray analysis (Section~\ref{sct:cont_xray}). The \textit{dbb} normalization equals to $R_{\rm in}^2~\cos i$, where $R_{\rm in}$ is the radius at the inner edge of the disk, and $i$ is the inclination angle of the disk. Here we assume $i$ equals to the reflection angle $24^{\circ}$ measured by \citet{Nardini2014}. Following \citet{Mehdipour2021}, with our best-fit \textit{dbb} parameters (Table~\ref{tbl:fit_02vs20sed}), we obtain $R_{\rm in}=0.10$~ld (or $\sim4.4~R_S$) and a mass accretion rate of $\sim2.5~M_{\odot}~{\rm yr^{-1}}$ for MR\,2251-178.


From our broadband SED modeling, the bolometric luminosity of MR\,2251-178 is $\sim1.71\times10^{45}~{\rm erg~s^{-1}}$ in 2002 and $\sim1.51\times10^{45}~{\rm erg~s^{-1}}$ in 2020. With the black hole mass of $(2.0\pm0.5)\times10^8~M_{\odot}$ \citep{Lira2011}, which is consistent with $\sim2.4\times10^8~M_{\odot}$ \citep{Dunn2008}, the Eddington ratio was $\sim6.8$~\% in 2002 and $\sim6.0$~\% in 2020. In 2011, using the XMM-Newton data (OM, RGS, and EPIC/pn), \citet{Nardini2014} derived a bolometric luminosity of $\sim(5-7)\times10^{45}~{\rm erg~s^{-1}}$, which corresponds to an Eddington ratio of $\sim20-28$~\%\footnote{\citet{Nardini2014} adopted a black hole mass of $2.4\times10^{8}~M_{\odot}$ \citep{Dunn2008}, which led them to report an Eddington ratio of $\sim15-25$~\%. }.

\begin{figure*}
\centering
\includegraphics[width=.83\hsize, trim={0.5cm 0.cm 2.cm 0.0cm}, clip]{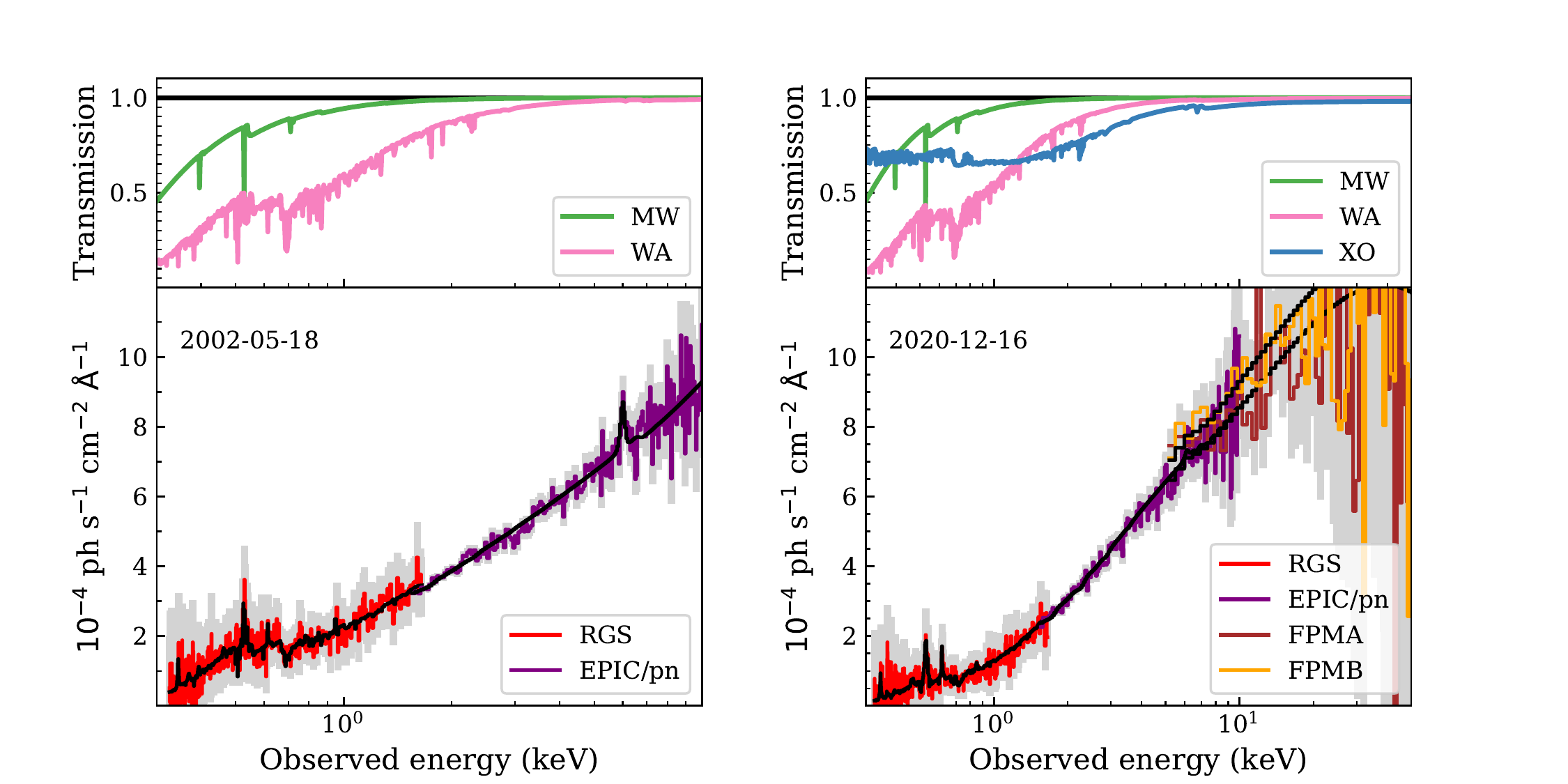}
\caption{Best-fit to the observed X-ray data of MR\,2251-178 in 2002 and 2020. Transmission of the Galactic absorption (MW), warm absorber (WA), and X-ray obscurer (XO, 2020 only) are shown in the top panels. In the bottom panels, data (colored curved with $1\sigma$ uncertainties in gray) and model (black curves) of each instrument are rebinned for clarity. The flux differences in the hard X-ray band are due to the cross-calibration issue between XMM-Newton and NuSTAR.  }
\label{fig:plot_dmtk_02vs20}
\end{figure*}

In Figure~\ref{fig:plot_dmtk_02vs20}, we show the best fit to the observed X-ray data, as well as the transmission of the Galactic absorption, warm absorber (X-ray), and obscurer (2020 only). The best-fit $C$-statistics \citep{Kaastra2017} are given in Table~\ref{tbl:fit_02vs20sed}, which includes the total $C$-statistics and statistics of individual instruments.

\subsection{X-ray warm absorber and emitter}
\label{sct:wawe}
The presence of the obscurer can significantly lower the soft X-ray flux so that the warm absorber features are less prominent. Therefore, the warm absorber parameters are better constrained with the X-ray spectrum obtained in 2002, where no obscurer was present along our line-of-sight. The best-fit parameters of the warm absorber are given in Table~\ref{tbl:fit_020518wa}. These parameters are consistent within the $1\sigma$ uncertainty of those reported by \citet{Reeves2013}.

The hydrogen column density ($N_{\rm H}$), microscopic turbulence velocity ($v_{\rm mic}$), and outflow velocity ($v_{\rm out}$) of the warm absorber derived from the 2002 X-ray spectrum are kept frozen when analyzing the 2020 X-ray spectrum. The warm absorber is expected to be less ionized in 2020 because photons from the central engine are screened by the obscurer before reaching the warm absorber. Since the warm absorber features are not detectable with the diminished soft X-ray flux, following our previous analysis on NGC\,5548 \citep{Kaastra2014} and NGC\,3783 \citep{Mehdipour2017,Mao2019}, we simply assume that the product $n_e~r^2$ of the warm absorber is constant, where $n_e$ is the number density and $r$ the distance of the warm absorber to the central engine. For the warm absorber in 2020, the ionization parameter \citep{Tarter1969,Krolik1981}
\begin{equation}
\label{eq:xi}
    \xi = \frac{L}{n_e~r^2}
\end{equation}
is then calculated with the $1-1000$ Rydberg ionizing luminosity. In 2020 (Table~\ref{tbl:fit_02vs20sed}), $\log \xi=$1.96, 0.79, and $-1.49$ for components \#$1-3$, respectively. Throughout this paper, the ionization parameter ($\xi$) is given in units of ${\rm erg~s^{-1}~cm}$.

\begin{table}[!b]
\centering
\caption{Best-fit parameters of the X-ray warm absorber (three components) in MR\,2251-178 observed on 2002-05-18. }
\label{tbl:fit_020518wa}
\begin{tabular}{l|c}
\hline\hline
\noalign{\smallskip}
Date & 2002-05-18 \\
\noalign{\smallskip}
\hline
\noalign{\smallskip}
\multicolumn{2}{c}{Warm absorber \#1} \\
\noalign{\smallskip}
\hline
\noalign{\smallskip}
$N_{\rm H}~({\rm 10^{21}~cm^{-2}})$ & $2.6_{-0.7}^{+1.4}$ \\
$\log \xi~{\rm (erg~s^{-1}~cm)}$ & $2.09_{-0.04}^{+0.10}$  \\
$v_{\rm mic}~{\rm (km~s^{-1})}$ & $46_{-13}^{+30}$ \\
$v_{\rm out}~{\rm (km~s^{-1})}$ & $-404_{-820}^{+170}$  \\
\noalign{\smallskip}
\hline
\noalign{\smallskip}
\multicolumn{2}{c}{Warm absorber \#2} \\
\noalign{\smallskip}
\hline
\noalign{\smallskip}
$N_{\rm H}~({\rm 10^{21}~cm^{-2}})$ & $2.43_{-0.04}^{+0.05}$ \\
$\log \xi~{\rm (erg~s^{-1}~cm)}$ & $0.92_{-0.10}^{+0.06}$  \\
$v_{\rm mic}~{\rm (km~s^{-1})}$ & $<30$ \\
$v_{\rm out}~{\rm (km~s^{-1})}$ & $-10_{-60}^{+270}$  \\
\noalign{\smallskip}
\hline
\noalign{\smallskip}
\multicolumn{2}{c}{Warm absorber \#3} \\
\noalign{\smallskip}
\hline
\noalign{\smallskip}
$N_{\rm H}~({\rm 10^{21}~cm^{-2}})$ & $1.40_{-0.03}^{+0.15}$  \\
$\log \xi~{\rm (erg~s^{-1}~cm)}$ & $-1.36_{-0.14}^{+0.26}$  \\
$v_{\rm mic}~{\rm (km~s^{-1})}$ & $<5$  \\
$v_{\rm out}~{\rm (km~s^{-1})}$ & $-370_{-180}^{+150}$ \\
\noalign{\smallskip}
\hline
\end{tabular}
\end{table}

\begin{figure}
\centering
\includegraphics[width=\hsize, trim={0.5cm 0.cm 0.5cm 0.0cm}, clip]{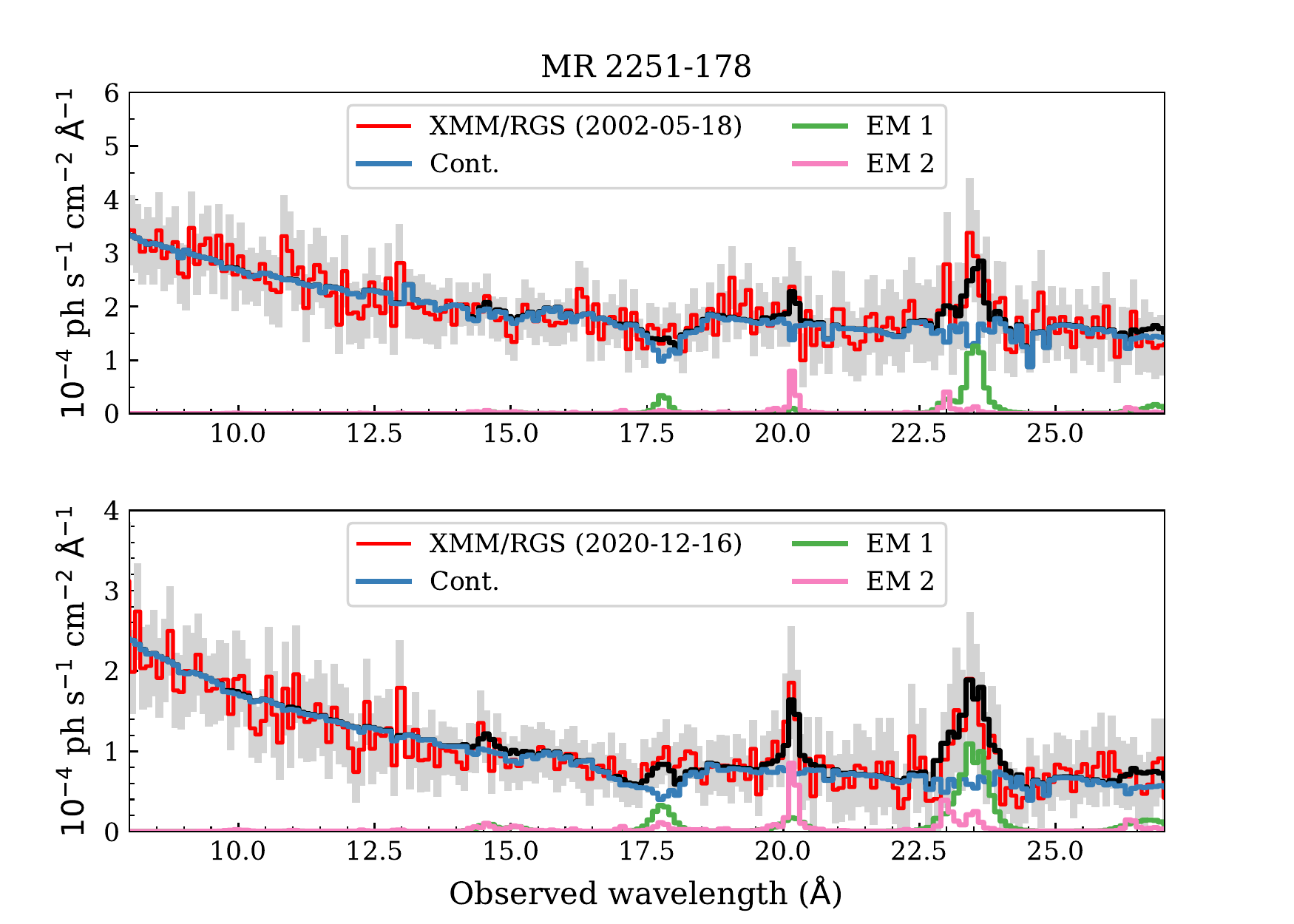}
\caption{The soft X-ray spectra of MR\,2251-178 in 2002 (top) and 2020 (bottom). The RGS spectra are rebinned for clarity. The black curves are the total flux, which consists of two warm emitter components (green and pink) and the continuum (blue).  }
\label{fig:plot_cf_dnma_02vs20}
\end{figure}

Two X-ray emission components with distinct velocity broadening are required for the soft X-ray emission lines in the 2002 and 2020 spectra (Figure~\ref{fig:plot_cf_dnma_02vs20} and Table~\ref{tbl:fit_02vs20em}). The warm emitter features are not prominent though, partly due to the relatively short exposure time. To reduce the number of free parameters, based on experience \citep[e.g.,][]{Mao2018,Mao2019,GWaters2021}, we fix the emission covering factor $C_{\rm em}=\Omega/4\pi$, where $\Omega$ is the solid angle subtended by the warm emitter with respect to the central engine. The emission covering factors of the two warm emitter components are 0.005 and 0.10, respectively. Both X-ray emission components are slightly (within $1-2\sigma$) broader and stronger in 2020. Nonetheless, the relatively short exposure for both X-ray spectra (Table~\ref{tbl:obs_log}) do not provide sufficient statistics to draw a firm conclusion.

\begin{table}[!t]
\centering
\caption{Best-fit parameters of the warm emitter (two components) in MR\,2251-178 observed in 2002 and 2020. Frozen parameters are indicated with (f). }
\label{tbl:fit_02vs20em}
\begin{tabular}{l|ll}
\hline\hline
\noalign{\smallskip}
Date & 2002-05-18 & 2020-12-16 \\
\noalign{\smallskip}
\hline
\noalign{\smallskip}
\multicolumn{3}{c}{Warm emitter \#1} \\
\noalign{\smallskip}
\hline
\noalign{\smallskip}
$N_{\rm H}~({\rm 10^{22}~cm^{-2}})$ & $10.7_{-1.6}^{+2.6}$ & $15.7_{-3.1}^{+3.4}$ \\
$\log \xi~{\rm (erg~s^{-1}~cm)}$ & $0.73_{-0.21}^{+0.14}$ & $0.97_{-0.14}^{+0.12}$ \\
$C_{\rm em}$ & 0.005 (f) & 0.005 (f) \\
$v_{\rm mac}~{\rm (km~s^{-1})}$ & $1330_{-260}^{+340}$ & $2480_{-530}^{+540}$ \\
\noalign{\smallskip}
\hline
\noalign{\smallskip}
\multicolumn{3}{c}{Warm emitter \#2} \\
\noalign{\smallskip}
\hline
\noalign{\smallskip}
$N_{\rm H}~({\rm 10^{20}~cm^{-2}})$ & $9.5_{-3.4}^{+10.7}$ & $20.1_{-8.3}^{+12.1}$ \\
$\log \xi~{\rm (erg~s^{-1}~cm)}$ & $1.36_{-0.11}^{+0.12}$ & $1.36$ (f) \\
$C_{\rm em}$ & 0.10 (f) & 0.10 (f) \\
$v_{\rm mac}~{\rm (km~s^{-1})}$ & $410_{-180}^{+230}$ & $630_{-210}^{+260}$ \\
\noalign{\smallskip}
\hline
\end{tabular}
\end{table}

\subsection{Obscurer}
\label{sct:ow}
The rising trend of the Swift X-ray hardness ratio since the summer of 2020 suggests an obscuration event, similar to those found in e.g., NGC\,5548 \citep{Kaastra2014} and NGC\,3783 \citep{Mehdipour2017}. The apparent lowering of the soft X-ray flux in 2020, in comparison with that in 2002, supports the presence of an obscurer. We aim to constrain the physical properties of the obscurer with X-ray, UV, and NIR data.

\subsubsection{X-ray spectra}
\begin{figure}
\centering
\includegraphics[width=.9\hsize, trim={0.5cm 0.cm 1.5cm 0.5cm}, clip]{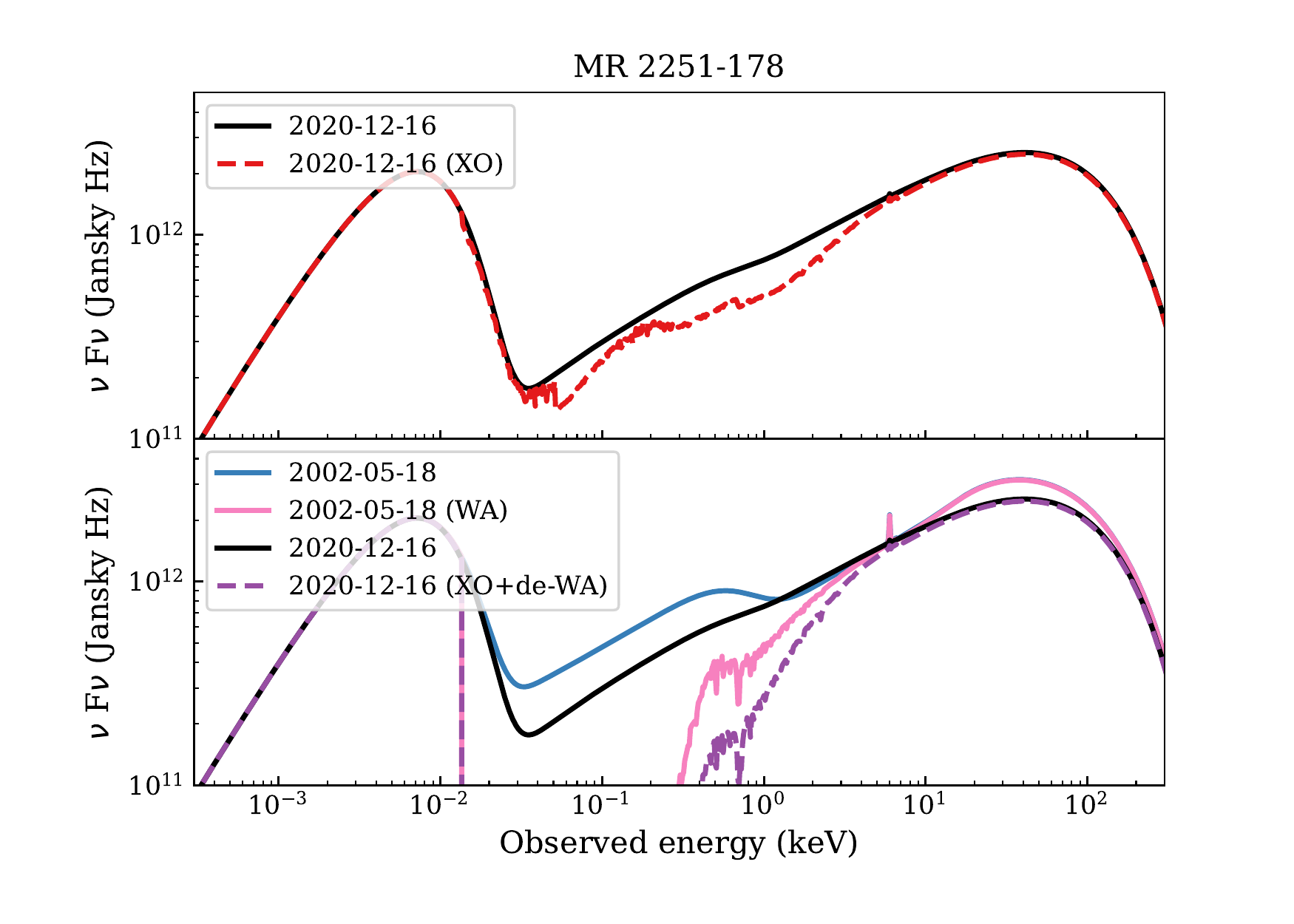}
\caption{The broadband spectral energy distribution (SED) of MR\,2251-178 with and without obscuration and absorption effects of ionized absorbers (Galactic absorption and extinction effects are excluded here). The upper panel shows the obscuration effect due to the X-ray obscurer (XO) in 2020. The pink curve in the lower panel includes the absorption effect by the X-ray warm absorber (WA) in 2002, while the purple curve includes both the obscuration effect by the obscurer and the absorption effect by the de-ionized warm absorber in 2020. }
\label{fig:plot_cf_sed_02vs20}
\end{figure}

The best-fit parameters of the obscurer are given in Table~\ref{tbl:fit_201216ow}. The hydrogen column density is $(8.2\pm1.6)\times10^{22}~{\rm cm^{-2}}$, which is comparable (within $1\sigma$) to the hydrogen column density $6.6_{-1.4}^{+0.8}\times10^{22}~{\rm cm^{-2}}$ reported by \citet{Markowitz2014}. The line-of-sight covering factor is $\sim0.35$. Hence, only a small fraction of the SED is obscured (Figure~\ref{fig:plot_cf_sed_02vs20}). That is to say, a large fraction of the X-ray photons leak through the obscurer. 

\begin{table}[!b]
\caption{Best-fit parameters of the obscurer in MR\,2251-178 observed on 2020-12-16. }
\label{tbl:fit_201216ow}
\centering
\begin{tabular}{l|c}
\hline\hline
\noalign{\smallskip}
Date & 2020-12-16 \\
\noalign{\smallskip}
\hline
\noalign{\smallskip}
\multicolumn{2}{c}{X-ray obscurer} \\
\noalign{\smallskip}
\hline
\noalign{\smallskip}
$N_{\rm H}~({\rm 10^{22}~cm^{-2}})$ & $8.2\pm1.6$ \\
$\log \xi~{\rm (erg~s^{-1}~cm)}$ & $1.83\pm0.13$ \\
$f_{\rm cov}^{X}$ & $0.35\pm0.02$ \\
\noalign{\smallskip}
\hline
\end{tabular}
\end{table}

\begin{figure}
\centering
\includegraphics[width=\hsize, trim={0.cm 0.cm 0.5cm 0.cm}, clip]{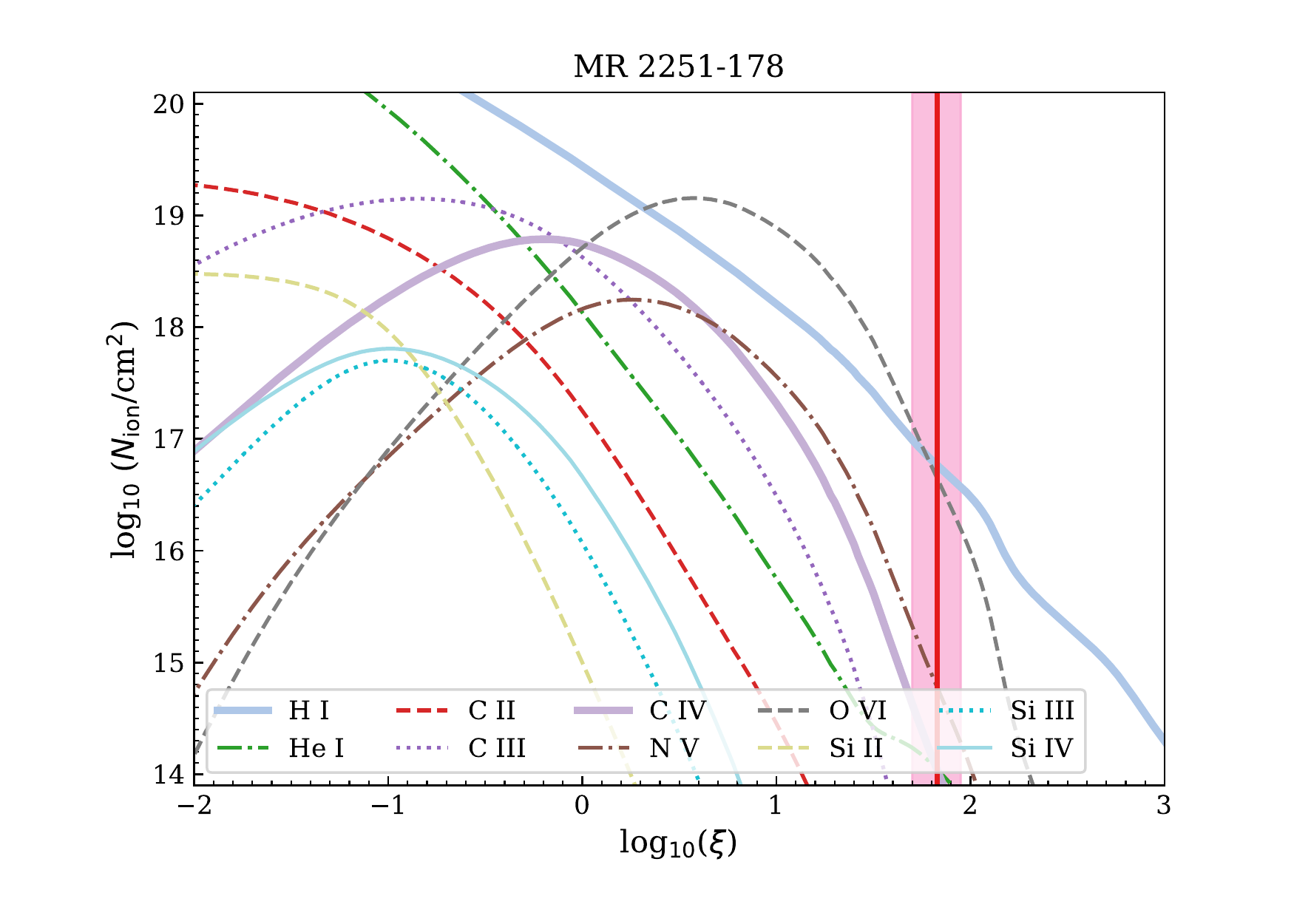}
\caption{Ionic column densities of the obscurer. This calculation is performed with the hydrogen column density and covering factor of the obscurer given in Table~\ref{tbl:fit_201216ow} and the incident ionizing SED shown in Figure~\ref{fig:plot_cf_sed_nir2xray}. The pink shaded area marks the X-ray constraints ($1\sigma$ uncertainty) of the ionization parameter.}
\label{fig:plot_cold_used_pion}
\end{figure}

With the incident ionizing SED (Figure~\ref{fig:plot_cf_sed_nir2xray}), and the hydrogen column density of the obscurer (Table~\ref{tbl:fit_201216ow}), the ionic column densities can be calculated (Figure~\ref{fig:plot_cold_used_pion}). For $\xi$ spanning several orders of magnitude, the corresponding ionic column densities of H {\sc i}, C {\sc iv}, O {\sc vi}, and so forth are well above $\sim10^{15}~{\rm cm^{-2}}$ . If the X-ray obscurer intercepts our line-of-sight to (a significant portion of) the UV emitting region, we expect to find absorption features in the HST/COS spectrum.


\subsubsection{UV line profiles}
\label{sct:uv_lpro}
In comparison with the archival HST spectra from $1996-2011$, new Ly$\alpha$ absorption features emerged in the 2020 HST/COS spectrum. These new Ly$\alpha$ absorption features are blueshifted by more than $2000~{\rm km~s^{-1}}$, well separated from the relatively slow warm absorber.

\begin{figure*}
\centering
\includegraphics[width=.83\hsize, trim={0.5cm 0.5cm 2.0cm 0.5cm}, clip]{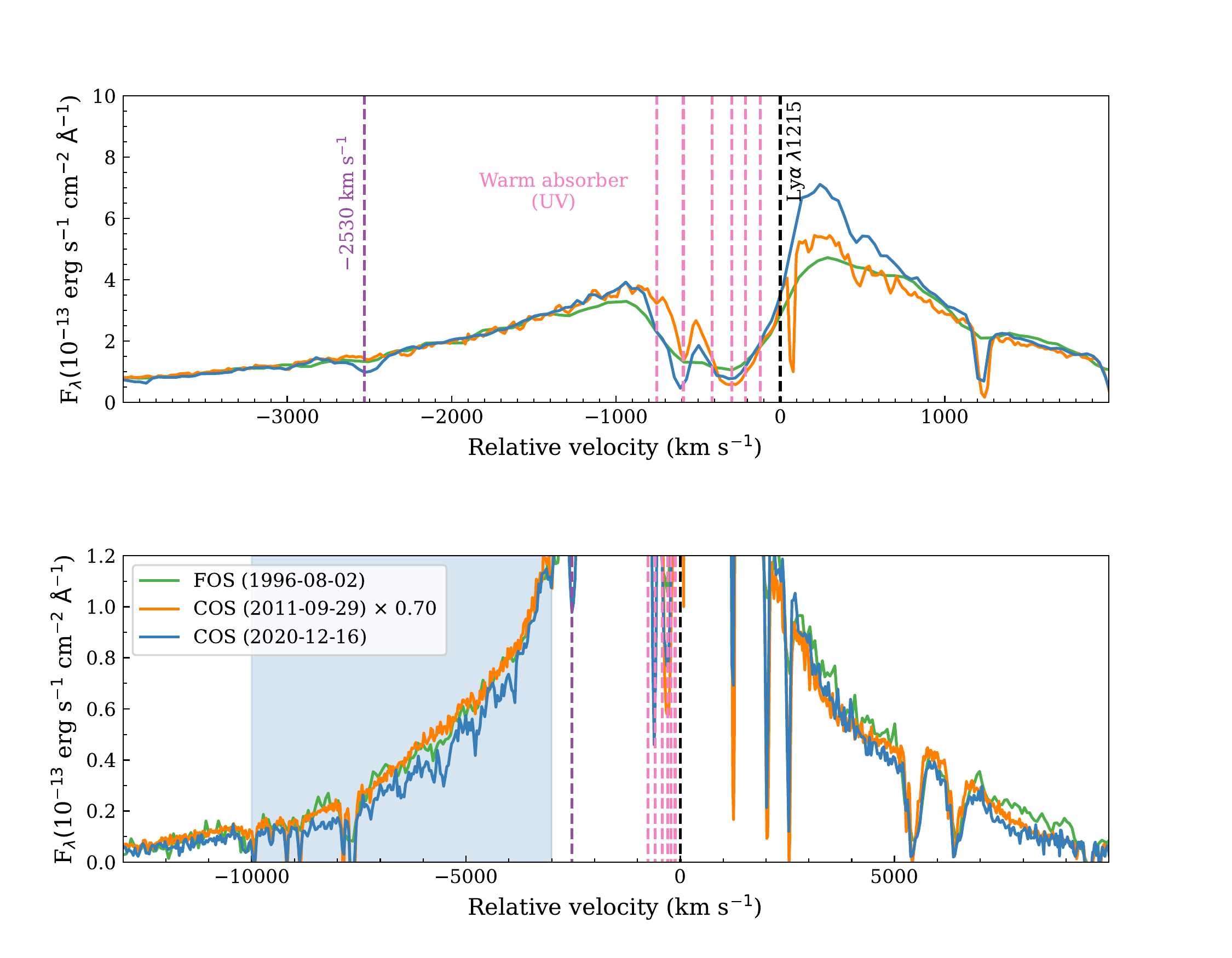}
\caption{The Ly$\alpha$ line profiles. The single-orbit 2020 spectrum is rebinned for clarity. Vertical dotted lines in pink mark the UV warm absorbers. Vertical dashed line in purple marks the new blueshifted ($\sim-2530~{\rm km~s^{-1}}$) narrow absorption line. A weak blueshifted broad absorption trough is observed between $\sim-3000~{\rm km~s^{-1}}$ and $\sim-10000~{\rm km~s^{-1}}$ accompanied by a forest of narrow absorption lines (the shaded area in light blue).  }
\label{fig:plot_cf_rel_uv_010101}
\end{figure*}

A new narrow absorption line was found in the blue wing of Ly$\alpha$ with an outflow velocity of $\sim-2530~{\rm km~s^{-1}}$ and a full-width-half-maximum (FWHM) of $\sim180~{\rm km~s^{-1}}$ (Figure~\ref{fig:plot_cf_rel_uv_010101}). This new absorption feature has an equivalent width of $3.1\pm0.3$~\AA. There might be a counterpart in the blue wing of O {\sc vi}, though the spectrum is noisy. No counterparts were found in lower ionization species (e.g., N {\sc v} and C {\sc iv}), which indicates that the ionization parameter of this UV absorber is relatively high.

\begin{figure}
\centering
\includegraphics[width=.9\hsize, trim={0.cm 0.cm 0.cm 0.cm}, clip]{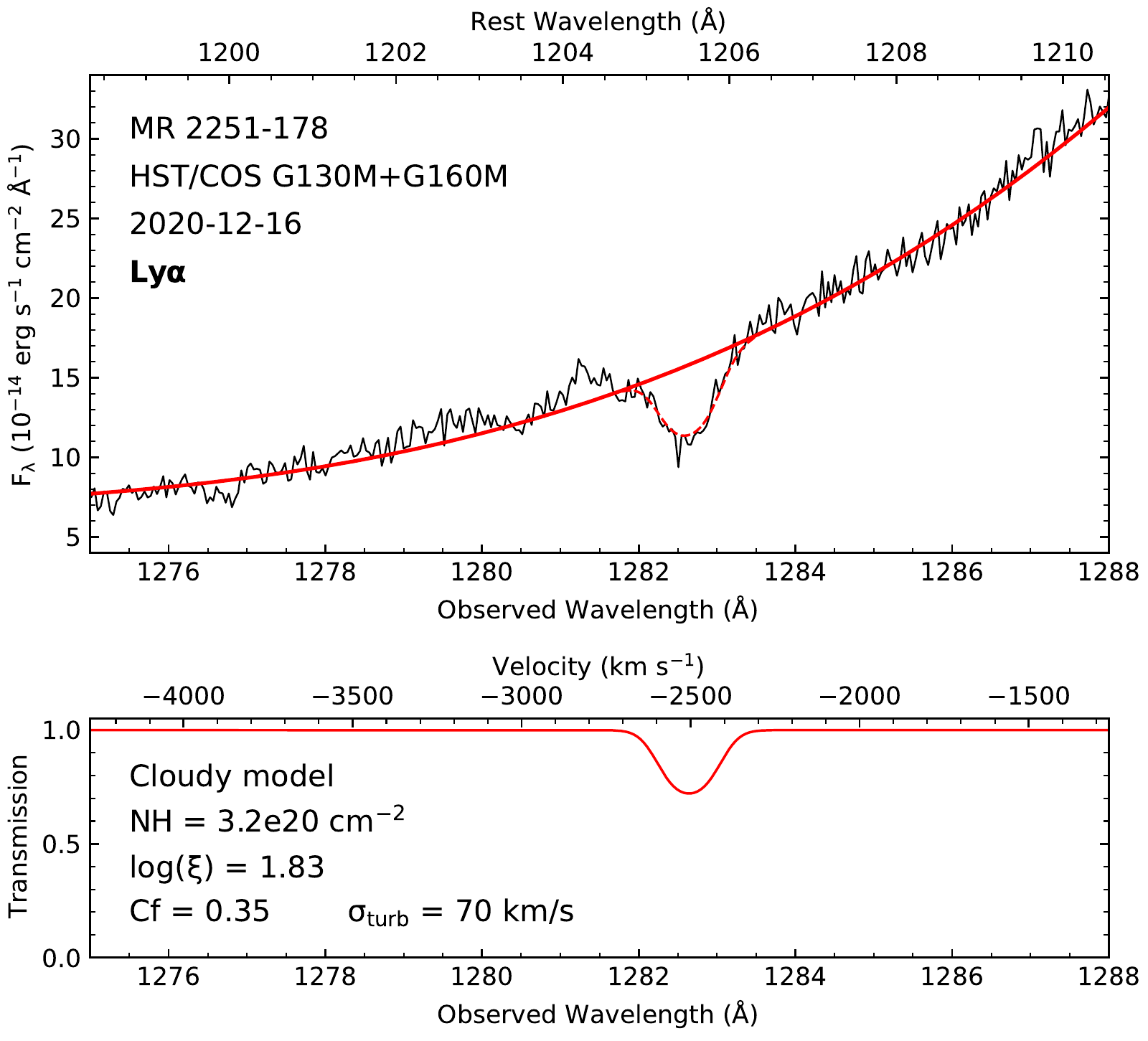}
\caption{Photoionization modeling of the new narrow absorption feature of Ly$\alpha$ using Cloudy \citep{Ferland2017}. With $\log \xi=1.8$ and $f_{\rm cov}^{\rm UV}=f_{\rm cov}^{\rm X}=0.35$, a good fit can be achieved with a hydrogen column density of $3.2\times10^{20}~{\rm cm^{-2}}$ and a turbulence velocity of $70~{\rm km~s^{-1}}$. The lower panel shows the transmission.}
\label{fig:plot_fit_lyalpha}
\end{figure}

Assuming $\log \xi=1.8$ and $f_{\rm cov}^{\rm UV}=f_{\rm cov}^{\rm X}=0.35$ (Table~\ref{tbl:fit_201216ow}), a good fit to the new narrow absorption line of Ly$\alpha$ can be achieved with $N_{\rm H}\sim3.2\times10^{20}~{\rm cm^{-2}}$ (Figure~\ref{fig:plot_fit_lyalpha}). In this case, turbulent broadening with $v_{\rm turb}\sim70~{\rm km~s^{-1}}$ is required, in addition to thermal broadening, to account for the observed broadening. The hydrogen column density measured in the UV band is more than two orders of magnitude lower than the X-ray measurement in Table~\ref{tbl:fit_201216ow}. 

\begin{figure}
\centering
\includegraphics[width=\hsize, trim={1.cm 1.0cm 1.5cm 1.0cm}, clip]{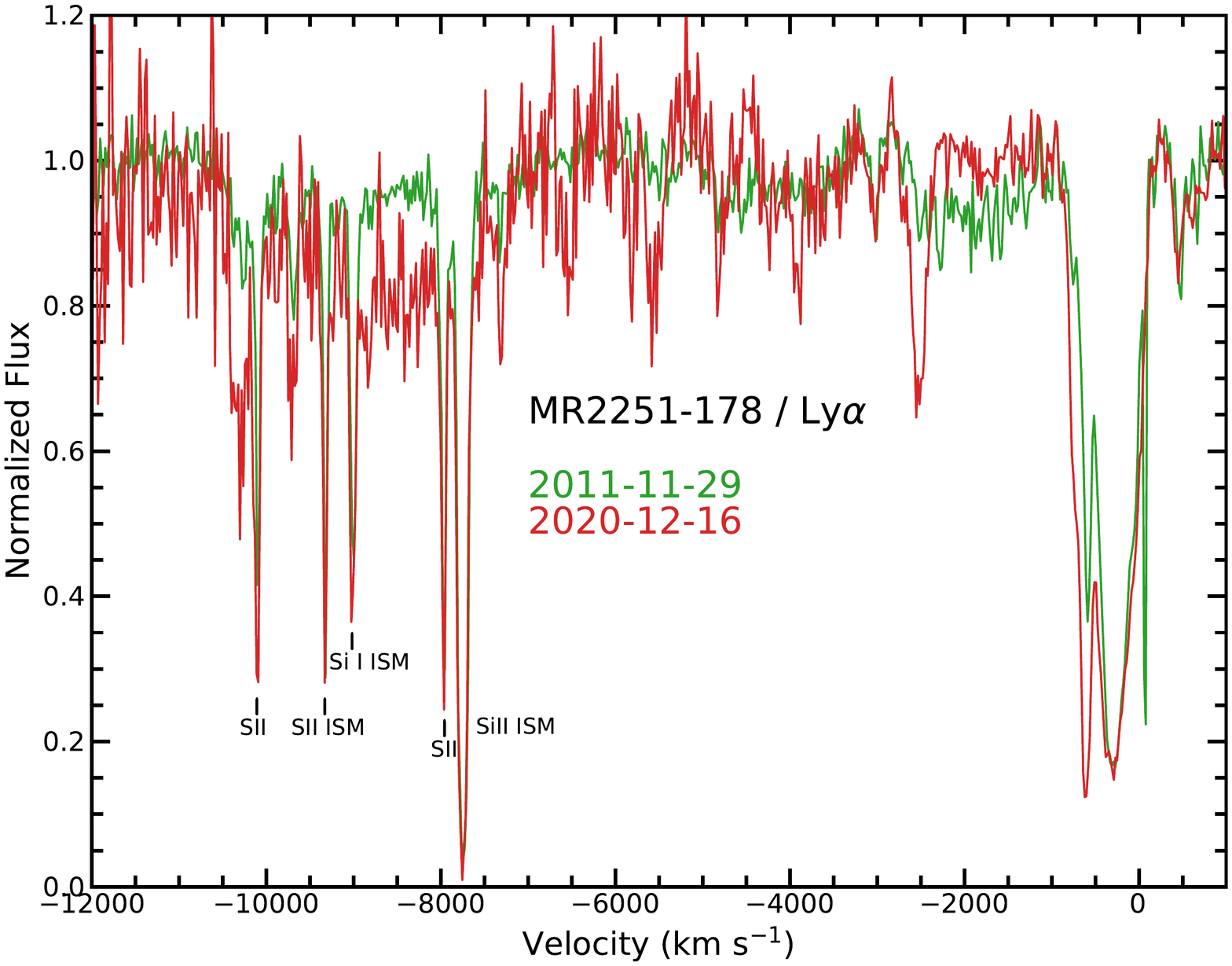}
\caption{Normalized flux of Ly$\alpha$ as a function of outflow velocity in 2011 (green) and 2020 (red) HST/COS spectra. In the 2020 spectrum, a forest of weak absorption features appear in the blue wing ranging from $\sim-2530~{\rm km~s^{-1}}$ to $\sim-10400~{\rm km~s^{-1}}$. }
\label{fig:plot_normflux_lyalpha}
\end{figure}

Moreover, a forest of weak absorption features appeared in the blue wing of Ly$\alpha$ in 2020, extending all the way down to $\sim-10400~{\rm km~s^{-1}}$ (Figure~\ref{fig:plot_normflux_lyalpha}). The apparent broad absorption feature between $\sim-8000~{\rm km~s^{-1}}$ and $\sim-9000~{\rm km~s^{-1}}$ might consist of multiple blended narrow absorption lines, if it is not intrinsically broad. No convincing counterparts are found in the blue wing of C {\sc iv}.

If the Ly$\alpha$ absorption features are all optically thin, then they cannot be the same gas responsible for the X-ray obscuration. Our fit to the strongest feature at $-2530~\rm km~s^{-1}$ shows that it is optically thin, which cannot be a counterpart to the X-ray obscuring gas. However, we have insufficient information to characterize the trough and forest extending to $-10400~\rm km~s^{-1}$. If these more extended features are optically thick and their weakness reflects low covering fractions, it is possible that this is the same gas as in the X-ray, but simply covering a much larger emitting area in the UV than for the compact X-ray source (as discussed in Sect.~\ref{sct:dist_uv} below). Additional possibilities for linking the UV and X-ray absorbing gas include a UV absorber with the same ionization and covering factor but a significantly lower column density, or UV absorbing gas with no parameters shared with the X-ray, and lying along a completely different line-of-sight.

In addition, a narrow absorption line in the red wing ($\sim+75~{\rm km~s^{-1}}$) of Ly$\alpha$ was present in the 2011-11-29 HST/COS spectrum but disappeared on 2020-12-16 (Fig.~\ref{fig:plot_cf_rel_uv_010101}). The relatively low velocity shift makes it less likely to be associated with the X-ray obscurer.

Variations of UV warm absorber features ($v_{\rm out}\lesssim10^3~{\rm km~s^{-1}}$) are found e.g., between $-600~{\rm km~s^{-1}}$ and $-800~{\rm km~s^{-1}}$) in the upper panel of Fig.~\ref{fig:plot_cf_rel_uv_010101}. These variations are mainly due to an ionization response of the lower (a factor of $\sim$3) flux of the UV and X-ray continuum in 2020 compared to 2011. The contribution from the lower ionizing luminosity caused by the obscurer is minor due to its relatively low covering factor (see also the upper panel of Fig.~\ref{fig:plot_cf_sed_02vs20}). This might be explained if our line-of-sights towards the X-ray and UV continuum intercept different parts of the obscuring wind.

\subsubsection{NIR and optical line profiles}
\label{sct:nir_opt_lpro}
\begin{figure}
\includegraphics[width=\hsize, trim={0.5cm 5.cm 1.5cm 1.cm}, clip]{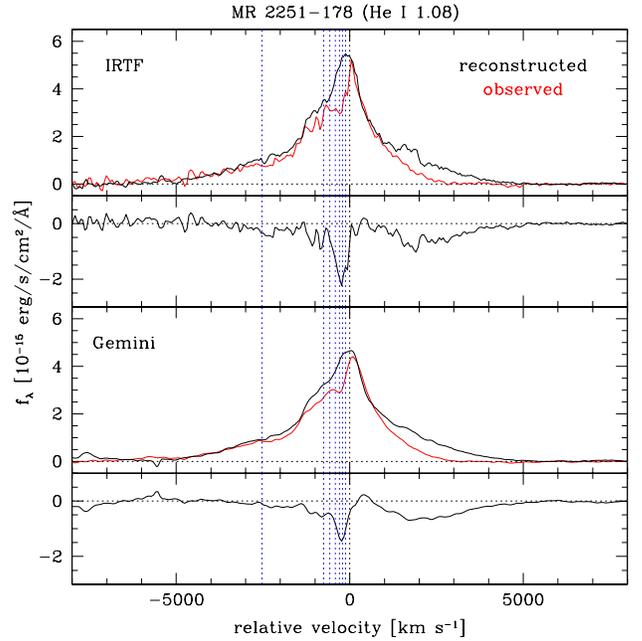}
\caption{Observed and reconstructed He {\sc i}* $\lambda10830$ line profile after removing the nearby Pa$\gamma$ $\lambda10938$ broad emission line. The dashed blue vertical lines mark the Ly$\alpha$ absorption components observed with HST/COS (upper left panel of Figure~\ref{fig:plot_cf_rel_uv_010101}). }
\label{fig:plot_cf_profile_NIR}
\end{figure}

Three NIR spectra of MR~2251-178 were collected in late 2020. While the Pa$\alpha$ line shows no absorption, clear, strong absorbers are evident in He~{\sc i}* $\lambda10830$. In Figure~\ref{fig:plot_cf_profile_NIR}, we show the isolated blueshifted He {\sc i}* absorption in the IRTF and Gemini NIR spectra and find them to be consistent with each other.

Since no previous NIR spectroscopy exists of MR~2251-178 from which we could extract an unabsorbed line profile, we reconstructed it using the strong Pa$\alpha$ broad line and [S {\sc iii}] $\lambda9531$ narrow line profiles from the same spectra. As \citet{Landt2008} showed, the Pa$\alpha$ and He~{\sc i}* broad line profiles are very similar. Furthermore, in MR~2251-178 we find that the Pa$\alpha$ line lacks a narrow component, making it ideal to model the broad component. Following the approach in \citet{Landt2008}, we also used the scaled Pa$\alpha$ line to remove the Pa$\gamma$ $10938$ emission line from the blend with He {\sc i}*.

We note that, in contrast to the broad-line AGN studied by \citet{Landt2008}, MR~2251-178 shows a deficit of broad line flux in the red wing of He~{\sc i}* line relative to the Pa$\alpha$ profile. This finding is confirmed when comparing the isolated He {\sc i}* $\lambda10830$ line with the He~{\sc i}* $\lambda3889$ line observed in the Keck optical spectrum, which is not blended with other ionic species on its red wing. The redshifted broad ($\sim 2200~{\rm km~s^{-1}}$) absorption trough is observed at a velocity of $+2235~{\rm km~s^{-1}}$. One of the possible interpretations is that this might be related to inflows fueling the accretion disk, as reported by \citet{Zhou2019} for eight other quasars. Another peculiarity that we find in MR~2251-178 is that all broad-line profiles are consistently blueshifted (by $\sim 300~{\rm km~s^{-1}}$) and that narrow emission lines from very low-ionisation gas, such as, e.g. \mbox{[O {\sc i}]}, \mbox{[N {\sc ii}]} and \mbox{[S {\sc ii}]}, are very weak and those from high-ionisation gas, such as, e.g. \mbox{[O {\sc iii}]}, \mbox{[Ne {\sc iii}]} and \mbox{[Ne {\sc v}]}, are relatively strong.

The new narrow absorption line seen in Ly$\alpha$ with $v_{\rm out}=-2530~{\rm km~s^{-1}}$ was not found in He {\sc i}*, as expected given the high ionization parameter and low column density of this absorber. We have deblended the IRTF spectrum, which has the best combination of spectral resolution and $S/N$, with four Gaussian components in the He {\sc i}* absorption; a broad feature at high velocity (width of $1443~{\rm km~s^{-1}}$, $v_{\rm out}=-2268~{\rm km~s^{-1}}$), a deep feature at low velocity (width of $417~{\rm km~s^{-1}}$, $v_{\rm out}=-252~{\rm km~s^{-1}}$) and two barely resolved features ($v_{\rm out}=-1099~{\rm km~s^{-1}}$ and $-823~{\rm km~s^{-1}}$). We note that given the intermediate spectral resolution of the NIR spectrum, all these features could be composed of several narrower features. It is noteworthy the similarity in velocity (and width) between the redshifted absorption trough and the fastest velocity outflow.

\begin{figure}
\includegraphics[width=\hsize, trim={0.5cm 5.cm 1.5cm 1.cm}, clip]{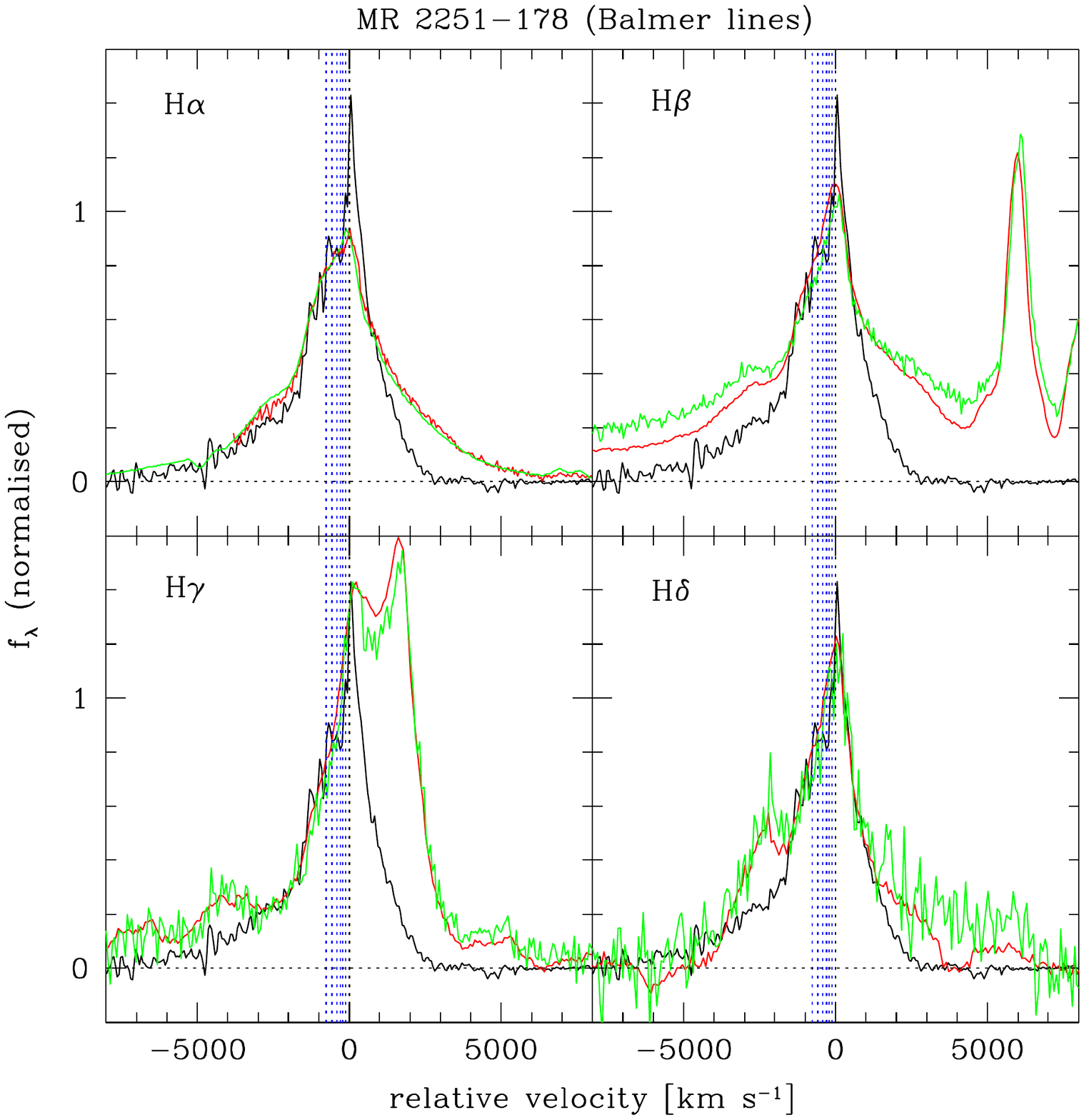}
\caption{Observed Balmer lines in the Keck (red) and Palomar (green) optical spectra compared to the observed He {\sc i}* $\lambda10830$ line profile (black) after removing the nearby Pa$\gamma$ $\lambda10938$ broad emission line. For H$\alpha$, we show the line profile from the IRTF spectrum instead since this strong line was saturated in the Keck spectrum. The dashed blue vertical lines mark the Ly$\alpha$ absorption components observed with HST/COS (upper left panel of Figure~\ref{fig:plot_cf_rel_uv_010101}). }
\label{fig:MR2251BalmerLines}
\end{figure}

It is tempting to associate the fastest velocity He {\sc i}* outflow with the X-ray obscurer itself, similar to what was observed in NGC~5548 \citep{Wildy2021}. Then, if this is indeed a wind off the accretion disk, as suggested by \citet{Dehghanian2019a, Dehghanian2019b}, the similarity in velocity (and width) between this feature and the redshifted absorption trough indicates that the geometry and viewing angle are such that we see also the far side, the receding part of this structure. This NIR absorber is likely co-located with the warm absorber and it must have only recently arisen due to a lower ionization parameter caused by the obscuration event. This situation is incompatible with the inflow scenario but would then be similar to what was recently observed in NGC\,5548 \citep{Wildy2021} and previously in NGC~4151 \citep{Hutchings2002, Wildy2016}. This interpretation is supported by the observed hydrogen Balmer line profiles (Figure~\ref{fig:MR2251BalmerLines}). The four strongest lines in this series show in both optical spectra an absorption profile similar to that of He~{\sc i}* $\lambda10830$. Notable is also the decreased absorption of the narrow line component as one goes up in the Balmer series (from H$\alpha$ to H$\delta$), indicating that the WA is co-located with that part of the narrow emission line region giving rise to permitted transitions and so of a higher density than the one further out and giving rise to forbidden transitions only.

\section{Discussion}
\label{sct:dis}
While there are more X-ray obscuration events in type 1 AGN in the literature \citep[e.g.,][]{Rivers2015, Gallo2021, Serafinelli2021, AAbendano2022}, we limit our discussions on the physical properties of the obscurer in quasar MR\,2251-178 in comparison with those observed with coordinated multi-wavelength observations. They are one radio-quiet quasar \citep[PG 2112+059,][]{Saez2021} and six Seyfert galaxies: NGC\,5548 \citep{Kaastra2014, Kriss2019b}, NGC\,985 \citep{Ebrero2016}, NGC\,3783 \citep{Mehdipour2017, Kaastra2018a, Kriss2019a}, Mrk\,335 \citep{Longinotti2013, Longinotti2019,Parker2019}, Mrk\,817 \citep{Miller2021, Kara2021}, and NGC\,3227 \citep{Mehdipour2017, YJWang2022, Mao2022}.

\subsection{Obscuration duration}
\label{sct:obs_duration}
The obscuration events in NGC\,5548 \citep{Kaastra2014,Mehdipour2016a,Mehdipour2022b} and Mrk\,335 \citep{Longinotti2019,Parker2019} lasted for years, while relatively shorter and recurrent obscuration events were found in NGC\,985 \citep{Ebrero2016}, NGC\,3783 \citep{Kaastra2018a}, Mrk\,817 \citep{Kara2021}, and NGC\,3227 \citep[e.g.,][]{Mehdipour2017, YJWang2022, Mao2022}. For short-lived obscuration events, the duration of the obscuration can be as short as 20~ks \citep[for NGC\,3227,][]{YJWang2022}.

For MR\,2251-178, if we consider the obscurer is present when the Swift X-ray hardness ratio is $\gtrsim0.20$ (Figure~\ref{fig:plot_uv_xhr}), the obscuration duration is $\sim75-260$~days. If the threshold is set to $\gtrsim0.25$, the obscuration duration would be $\sim75$ days.

\subsection{Obscuration recurrence}
\label{sct:obs_cycle}
Short-lived obscuration events are found to be recurrent for NGC\,3783, NGC\,985, and NGC\,3227. For NGC\,3783, the Swift hardness ratios between 2008 and 2017 suggest that there were obscuration events in early 2009 \citep{Kaastra2018a} and December 2016 \citep[][with joint UV and X-ray observations]{Mehdipour2017}. Furthermore, archival X-ray spectra obtained with the Advanced Satellite for Cosmology and Astrophysics (ASCA) in 1993 and 1996 also show evidence of obscuration events \citep{Kaastra2018a}. For NGC\,985, an obscurer was present with joint X-ray and UV observations in August 2013 and January 2015 \citep{Ebrero2016}. Archival X-ray data suggest that there might have been another obscuration event in July 2003 \citep{Ebrero2016}. For Mrk\,817, obscuration has been persistent and variable from November 2020 through March 2021 \citep{Kara2021}, including a 55-day period similar to the ``broad-line holiday" observed in NGC\,5548. Moreover, an archival NuSTAR spectrum taken on 2015-07-25 suggests an earlier obscuration event, but less prominent than the late 2020 event \citep{Kara2021}. For NGC\,3227, X-ray obscuration events were reported in $2000-2001$ \citep{Lamer2003,Markowitz2014}, 2002 \citep{Markowitz2014}, 2006 \citep{YJWang2022}, 2008 \citep{Beuchert2015}, 2016 \citep{Turner2018, YJWang2022}, and 2019 \citep{Mao2022}.

For PG 2112+059, obscuration events were observed on 2014-12-20 and 2015-08-29, but not on 2002-09-01 \citep{Saez2021}. With no observations between 2014-12-20 and 2015-08-29, it is hard to tell whether it was a continuous obscuration event lasting for eight months or two short-lived events.

For MR\,2251-178, X-ray obscuration events might have occurred in 1980 \citep{Halpern1984}, 1996 \citep{Markowitz2014}, and late 2020 (present work). Compared to the Einstein X-ray spectrum of MR 2215-178 taken on 1979-07-01, the one on 1980-05-19 was significantly absorbed in the soft X-ray band below $\sim3$~keV \citep{Halpern1984}. \citet{Markowitz2014} identified a clear obscuration event in the RXTE observation on 1996-12-09. In the two BeppoSAX observations in June and November 1998, X-ray obscuration had disappeared \citep{Dadina2007}. No other obscuration events were identified with the RXTE data up to January 2012 \citep{Markowitz2014} and Swift data between 2014 and mid-2020 (Figure~\ref{fig:plot_ltc_uvx}). With these X-ray observations and identified obscuration events, if there is a duty cycle, it would range from 16 to 24 years. Some periodic events might trigger the launch of the obscuring wind, or a warped accretion disk might cross our line-of-sight.


\subsection{Distance of the obscurer to the black hole}
It is not trivial to estimate the distance of the obscurer to the black hole since we cannot directly resolve it via imaging. \citet{Lamer2003} estimated the distance of the obscurer in NGC\,3227 assuming it is a spherical cloud orbiting the black hole in a Keplerian orbit. To be more specific, the distance is derived by (1) assuming the radius of the obscurer is $N_{\rm H}/n_{\rm H}$; (2) the obscurer crosses our line-of-sight with a constant velocity $v_{\rm cross}=\sqrt{GM_{\rm BH}/r}$ in a Keplerian orbit; (3) the crossing time $t_{\rm cross}=N_{\rm H}/(n_{\rm H}~v_{\rm cross})$ is constrained from the observed duration of the event. We estimate the distance of the 2020 obscurer in MR\,2251-178 following the above approach but also discuss alternative scenarios below.

\subsubsection{Spherical clouds in Keplerian orbits}
\label{sct:dist2BH_orbit}
For the 1996 obscuration event (with a duration of $3-1641$~days) of MR\,2251-178, \citet{Markowitz2014} estimated its distance to the black hole to be $\sim460-5700$~ld, implying a spherical obscurer with a size scale of $0.009-3.4$~ld. They also estimated the H$\beta$ distance ($\sim27$~ld) and the boundary of the dust sublimation zone \citep{Nenkova2008} $r_d\sim910$~ld. For $r\gtrsim r_d$, dust likely does not sublimate, while regions with $r\lesssim(0.3-0.5)~r_d$ are dust-free. Thus, they expect the obscurer in 1996 to be dusty.

For the 2020 event, following the same approach, we would obtain $r\sim155-255$~ld or $7000-11000~r_S$ and a size scale of $(0.7-1.8)$~ld. According to \citet{Nenkova2008}, with $L_{\rm bol}\sim1.50\times10^{45}~{\rm erg~s^{-1}}$ (Section~\ref{sct:sed}) and a dust temperature $T_d=1500$~K, the boundary of the dust sublimation zone is $\sim583$~ld. We can also derive the luminosity-based dust radii following \citet{Landt2019}, which yields $r_{d}\sim166$~ld for MR\,2251-178. These estimations suggest that the obscurer in 2020 is also likely dusty. With Equation~\ref{eq:xi}, the number density of the obscurer is $n_{\rm H}\sim(1.8-4.8)\times10^7~{\rm cm^{-3}}$, which is significantly smaller than the typical number density ($\sim10^{9-13}~{\rm cm^{-3}}$) of BLR clouds \citep{Peterson2006}.

\subsubsection{Non-spherical clouds in Keplerian orbits}
\label{sct:dist2BH_fscale}
It is possible that the obscurer is not spherical in shape.
As discussed in \citet{Mao2022}, based on the distance estimate for spherical clouds in Keplerian orbits, we can correct for the non-spherical geometry effect by introducing the azimuthal to radial size ratio $f$. With $f=1$, we obtain the distance estimation equations used by \citet{Lamer2003} and \citet{Markowitz2014}. For larger $f$ values, the estimated distance would be closer to the black hole (scaled by $f^{-2/5}$). Since the geometrical shape of the obscurer is unknown, the distance estimates assuming $f=1$ might be under- or over-estimated by more than an order of magnitude if $f\lesssim0.004$ or $f\gtrsim300$.

\subsubsection{Outflows launched from Keplerian orbits}
\label{sct:dist2BH_wind}
Here, we estimate the distance that does not rely on the unknown geometrical shape of the obscurer. In the above scenarios (Sect.~\ref{sct:dist2BH_orbit} and Sect.~\ref{sct:dist2BH_fscale}), the orbiting clouds are expected to have zero velocity in the radial direction. The observed Ly$\alpha$ absorption features are blueshifted with outflow velocities ranging from $-2530~{\rm km~s^{-1}}$ to $-10040~{\rm km~s^{-1}}$ (Fig.~\ref{fig:plot_normflux_lyalpha}), indicating that the obscurer is outflowing in the radial direction.

Given the relatively low outflow velocity in the radial direction (Fig.~\ref{fig:plot_normflux_lyalpha}) and short traveling time (Sect.~\ref{sct:obs_duration}), the obscurer is expected to be close to its launching radius. Assuming the obscurer was launched from a Keplerian orbit with a period of $P_{\rm orb}\sim16-24$~yr around the black hole (Sect.~\ref{sct:obs_cycle}), the distance of the obscurer is then $r=(GM_{\rm BH}P_{\rm orb}^2/4\pi^2)^{1/3}\sim21.4-28.1$ ld (or $\sim940-1200~r_S$). This is comparable to the H$\beta$ distance ($\sim27$~ld) given in \citet{Markowitz2014}. This is also in accordance with the line-driven disk winds launched at a few $\times10^2~r_S$ radii in numerical simulations \citep[e.g.,][]{Nomura2013, Mizumoto2021}. We can further estimate the number density of the obscurer via Eq.~\ref{eq:xi}, which yields $n_{\rm H}\sim(0.5-7.4)\times10^{11}~{\rm cm^{-3}}$. This value is within the range ($\sim10^{9-13}~{\rm cm^{-3}}$) of typical number density of BLR clouds \citep{Peterson2006}.



\subsection{X-ray and UV (dis)connection}
\label{sct:dist_uv}
During the X-ray obscuration period of NGC\,5548 \citep{Kaastra2014,Mehdipour2016a}, NGC\,985 \citep{Ebrero2016}, NGC\,3783 \citep{Mehdipour2017, Kaastra2018a}, Mrk\,335 \citep{Longinotti2019, Parker2019}, and Mrk\,817 \citep{Kara2021}, prominent blueshifted broad absorption troughs were found in the simultaneous HST/COS spectra. For the best-studied obscurer in NGC\,5548, both its column density and covering fraction are variable on timescales of a few ks to a few months \citep{DiGesu2015, Mehdipour2016a, Cappi2016}. The rapid variability and large velocity broadening support the scenario that the obscurer in NGC\,5548 originates from the accretion disk \citep{Kaastra2014}. For NGC\,3783, an additional high-ionization absorption component was also present in late 2016, leading to Fe {\sc xxv} and Fe {\sc xxvi} absorption lines in the Fe K band \citep{Mehdipour2017}. Through a detailed UV analysis, \citet{Kriss2019a} suggest that a collapse of the broad-line region (BLR) clouds triggered the launch of the obscurer. For Mrk\,817, the obscurer is located at the inner BLR and partially covers the central source \citep{Kara2021}.

For NGC\,3227, we do not find prominent blueshifted broad absorption troughs in the simultaneous HST/COS spectra when the target is obscured in the X-ray band \citep{Mao2022}. The lack of X-ray and UV association might be explained if the X-ray obscurer does not intercept our line-of-sight to (a significant portion of) the UV emitting region. If a compact X-ray obscurer intercepts our line-of-sight to the UV emitting region, it might cover $\lesssim1$~\% of the UV emitting region. It is also possible that the X-ray obscurer does not intercept our line-of-sight to the UV emitting region at all, although we cannot well constrain the distance of the obscurer.

For MR\,2251-178, following \citet[][Eq. S7]{Burke2021}, the effective UV emitting region radius ($R_{2500}$) can be estimated via
\begin{equation}
    R_{2500} = 10^{14.95\pm0.05}~{\rm cm}~\left(\frac{L_{5100}}{10^{44}~{\rm erg~s^{-1}}}\right)^{0.53\pm0.04}~,
\end{equation}
where $L_{5100}$ is the optical continuum luminosity. With $L_{5100}\sim1.4\times10^{44}~{\rm erg~s^{-1}}$, we obtain $R_{\rm 2500}\sim1.1\times10^{15}~{\rm cm}$ or $\sim0.41$~ld. This is equivalent to $\sim18~R_{\rm S}$, where the Schwarzschild radius $R_{\rm S}=2GM_{\rm BH}/c^2=5.9\times10^{13}~{\rm cm}$ or $0.023$~ld.

Assuming a fiducial X-ray emitting central engine with a size scale of $\lesssim5~R_{\rm S}$ \citep[e.g.,][]{Reis2013, Fabian2015}, the UV emitting region would be a factor of $\gtrsim13$ larger. With the X-ray obscurer covering $\sim35$~\% of the X-ray emitting central engine (Table~\ref{tbl:fit_201216ow}), it would then cover merely $\lesssim2.7$~\% of the UV emitting region.

The X-ray covering factor ($f_{\rm cov}^{X}$) of MR\,2251-178 is comparable to that reported by \citet{Ebrero2016} for the observation episode in NGC\,985 observed with XMM-Newton in January 2015. In that case, only one narrow (FHWM$\sim350\pm50~{\rm km~s^{-1}}$) blueshifted ($-5300\pm10~{\rm km~s^{-1}}$) absorption line of Ly$\alpha$ was present in the simultaneous HST/COS spectrum. The UV covering factors for the low  er-ionization ions C {\sc iii}*, C {\sc iv}, and Si {\sc iv} were $\lesssim0.15$ and not visible at the signal to noise ratio of the UV spectra. 

On one hand, if the obscuring wind barely intercepts the UV continuum region (e.g., the wind projected size is too small compared to the UV emitting region), we might not observe prominent blueshifted absorption lines in the UV band. On the other hand, when a transient obscuring wind is fading, we might not observe prominent blueshifted absorption lines in the UV band. Both might contribute to the lack of prominent UV absorption lines in MR\,2251-178.

\section{Summary}
\label{sct:sum}
We present the recent multi-wavelength observations of MR\,2251-178, including a set of Swift snapshots in $2020-2021$, a coordinated observation with HST/COS, XMM-Newton, and NuSTAR on 2020-12-16, and NIR to optical spectroscopic observations using NASA's Infrared Telescope Facility, Gemini, Keck, and Palomar from September to December 2020.

We construct a broadband spectral energy distribution of the AGN continuum with the multi-wavelength data. We perform a spectral analysis of the X-ray data, including the photoionization modeling of the newly discovered obscurer and the classical warm absorber in the X-ray band. The observational features in the X-ray band shared similarities with some other type 1 AGNs with obscuring winds. The observational characteristics of the spectral features associated with the obscuring wind in the UV to NIR bands are distinctly different from those seen in other AGN with obscuring outflows. A general understanding of the observational variety of obscuring winds in type 1 AGNs is still lacking. The main results are summarized as follows.

\begin{enumerate}
    \item In the X-ray band, an obscuration event was present in late 2020, which led to a flux drop in the soft X-ray band without any detectable absorption line features. The obscurer has a hydrogen column density of $\sim8.2\times10^{22}~{\rm cm^{-2}}$, a relatively high ionization parameter ($\log \xi\sim1.8$), and a relatively low line-of-sight covering factor ($f_{\rm cov}^{X}\sim0.35$).

    \item In the UV band, a new blueshifted narrow absorption line of Ly$\alpha$ with an outflow velocity of $\sim-2530~{\rm km~s^{-1}}$ was found in 2020. This new feature is accompanied by a forest of weak absorption features extending to $\sim-10400~{\rm km~s^{-1}}$ in the blue wing of Ly$\alpha$. No such absorption features are found in the blue wing of C {\sc iv}. The new weak UV absorption features can only be produced by the same gas obscuring the X-ray if it has a much lower covering fraction relative to the size of the UV-emitting regions.

    \item In the NIR band, weak narrow absorption lines are present in the blue wing of He {\sc i} 1.08~$\mu$m down to $\sim-2268~{\rm km~s^{-1}}$. Although no previous NIR spectroscopy exists for MR\,2251-178, we expect the de-ionized warm absorber gives rise to the narrow absorption lines. A redshifted broad absorption trough peaking around $+2235~{\rm km~s^{-1}}$ with a width of $\sim2200~{\rm km~s^{-1}}$ was found for He {\sc i} $\lambda10830$. It might be due to an inflow fueling the accretion disk. Alternatively, considering absorption features on both sides of He {\sc i}, it is possible that we are observing the far (receding) side of the obscurer.

    \item According to Swift hardness ratios in $2020-2021$, the transient obscurer in MR\,2251-178 might last for $75-260$ days. X-ray obscuration events might have a quasi-period of two decades considering earlier events in 1980 and 1996.

    \item It is hard to conclude the exact location of the 2020 X-ray obscurer. The inferred distance ranges from the BLR to the dusty torus. However, it is expected to be far beyond the UV emitting region.
\end{enumerate}

\section*{Acknowledgements}
We thank the referee for providing useful feedback to improve the quality of this work. A reproduction package is available at Zenodo \href{https://sandbox.zenodo.org/record/1065561#.YzKu3OxByWA}{DOI: 10.5072/zenodo.1107961}. This package includes data and scripts used to reproduce the fitting results and figures presented here. This work is based on observations obtained with XMM-Newton, an ESA science mission with instruments and contributions directly funded by ESA Member States and the USA (NASA). This work was supported by NASA through a grant for HST program number 16465 from the Space Telescope Science Institute, which is operated by the Association of Universities for Research in Astronomy, Incorporated, under NASA contract NAS5-26555. This research was supported under NASA Contract No. NNG08FD60C, and has made use of data obtained with the NuSTAR mission, a project led by the California Institute of Technology (Caltech), managed by the Jet Propulsion Laboratory (JPL) and funded by NASA. We thank the Swift Science Operations team for approving Swift target-of-opportunity observations in August and December 2020, as well as May 2021. We thank Claus Leitherer for approving the Director's Discretionary program, Nick Indriolo and Tricia Royle for technical support and executing the HST observation. We thank Norbert Schartel for approving the XMM-Newton directory's discretionary time observation. We thank Fiona Harrison for approving the NuSTAR Target of Opportunity. We also thank the HST, XMM, and NuSTAR teams for coordinating the observations. SRON is supported financially by NWO, the Netherlands Organization for Scientific Research. JM acknowledges useful discussions with M. S. Brotherton, Pu Du, Dongwei Bao, Zhu Liu, and Shasha Li. HL acknowledges a Daphne Jackson Fellowship sponsored by the Science and Technology Facilities Council (STFC), UK. MMK acknowledges generous support from the David and Lucille Packard Foundation.


%

\vspace{5mm}
\facilities{IRTF(SpeX), Gemini(GNIRS), Keck(LRIS), Hale(TSpec and DBSP), HST(COS and FOS), FUSE, Swift(XRT and UVOT), XMM, NuSTAR}


\software{
astropy \citep{Astropy2013},
SPEX v3.05.00 \citep{Kaastra1996,Kaastra2018b},
Cloudy v17 \citep{Ferland2017},
}

\bibliographystyle{aasjournal}
\bibliography{refs}



\end{document}